\documentclass[fleqn,usenatbib]{mnras}

\usepackage{mathptmx}

\usepackage[T1]{fontenc}

\DeclareRobustCommand{\VAN}[3]{#2}
\let\VANthebibliography\thebibliography
\def\thebibliography{\DeclareRobustCommand{\VAN}[3]{##3}\VANthebibliography}

\usepackage{graphicx}	%
\usepackage{amsmath}	%
\usepackage[dvipsnames]{xcolor}

\title[Young stellar populations in TIMER galaxies]{Model-dependent metallicities of young stellar populations in TIMER galaxies}

\author[G. Gonçalves et al.]{
Geraldo Gonçalves$^{1,2}$, 
Paula R. T. Coelho$^{1}$, 
Luiz A. Silva-Lima$^{3,4}$, 
Lucimara Martins$^{3}$, 
Marie Martig$^{2}$\thanks{E-mail: M.Martig@ljmu.ac.uk}, \newauthor
Dimitri A. Gadotti$^{5}$, 
Camila de Sá-Freitas$^{6}$, 
Jairo Mendez-Abreu$^{7,8}$,
Patricia Sánchez-Blázquez$^{9}$, \newauthor
Jesús Falcón-Barroso$^{7,8}$,
Glenn van de Ven$^{10}$,
Adriana de Lorenzo-Cáceres$^{7,8}$,
Isabel Pérez$^{11,12}$
\\
$^{1}$Instituto de Astronomia, Geofísica e Ciências Atmosféricas, Universidade de São Paulo, Rua do Matão 1226, Cidade Universitária, 05508-090 São Paulo, SP, Brazil\\
$^{2}$Astrophysics Research Institute, Liverpool John Moores University, 146 Brownlow Hill, Liverpool L3 5RF, UK\\
$^{3}$NAT–Universidade Cidade de São Paulo, Rua Galvão Bueno, 868, 01506-000, Sao Paulo, SP, Brazil\\
$^{4}$Laboratório Nacional de Astrofísica (LNA), Rua dos Estados Unidos 154, Bairro das Nações, 37504-364, Itajubá, MG, Brazil.\\
$^{5}$Centre for Extragalactic Astronomy, Department of Physics, Durham University, South Road, Durham DH1 3LE, UK\\
$^{6}$European Southern Observatory, Alonso de Córdova 3107, Vitacura, Casilla 19001, Santiago, Chile\\
$^{7}$Universidad de La Laguna, Departamento de Astrofísica, Avenida Astrofísico Francisco Sánchez s/n, 38205 La Laguna, Tenerife, Spain\\
$^{8}$Instituto de Astrofísica de Canarias, Calle Vía Láctea s/n, 38200 La Laguna, Tenerife, Spain\\
$^{9}$Departamento de Física de la Tierra y Astrofísica, Universidad Complutense de Madrid, E-28040, Madrid, Spain\\
$^{10}$European Southern Observatory, Karl-Schwarzschild-Str. 2, D-85748 Garching bei München, Germany\\
$^{11}$Departamento de Física Teórica y del Cosmos, Facultad de Ciencias (Edificio Mecenas), Universidad de Granada, E-18071 Granada, Spain\\
$^{12}$Instituto Carlos I de Física Teórica y Computacional, Universidad de Granada, E-18071 Granada, Spain
}

\date{Accepted XXX. Received YYY; in original form ZZZ}

\pubyear{\the\year{}}

\begin{document}
\label{firstpage}
\pagerange{\pageref{firstpage}--\pageref{lastpage}}
\maketitle

\begin{abstract}
Stellar populations in the central regions of disc galaxies provide key constraints on their formation and secular evolution.
We investigate the properties of young stellar populations in the centres of nearby barred galaxies from the TIMER survey, using four stellar population models, two based on empirical stellar libraries (MILES, X-Shooter) and two on theoretical libraries (Starburst99, HR-pyPopStar), complemented by independent gas-phase metallicities from emission-line diagnostics.
The inferred stellar metallicities depend strongly on the adopted model. 
Measured as the 5th-to-95th percentile spread across the three galaxies analysed in detail, the empirical libraries recover variations of $1.11$ and $0.98$\,dex, whereas the theoretical ones return $0.40$ and $0.49$\,dex; the gas in the same regions spans $0.49$\,dex.
Gas metallicities are indeed lower in the central star-forming regions, but the amplitude of that depression (about $0.3$\,dex in NGC\,1097) exceeds what bar-driven inflow of disc gas can account for, given typical abundance gradients in literature.
The age--metallicity relations consistently reveal two distinct components: an older population consistent with a self-enrichment scenario, and a younger population formed from gas inflows with a different chemical enrichment history. 
The empirical libraries produce a monotonic decline of metallicities from old to young populations, which would require a sustained inflow of metal-poor gas outpacing its own enrichment. The theoretical libraries instead recover a young component whose metallicity departs from the older populations only at the most recent ages and by a few tenths of a dex, compatible with a single dilution episode.
Our results demonstrate that model choice critically affects inferred metallicities, with models based on empirical stellar libraries recovering age--metallicity relations that demand physical conditions the gas measurements cannot sustain.
\end{abstract}


\section{Introduction}

Observations of extragalactic systems at low redshift show that massive disc galaxies generally display ordered kinematics and relatively smooth rotation curves. In contrast, disc galaxies observed at high redshift are significantly more turbulent, clumpy, and morphologically irregular \citep{forster+06, law+09}. Recent studies based on \textit{JWST} observations suggest that this transition may occur at redshifts around $z \sim 4$ (\citealt{ferreira+23}, \citealt{company+25}). Understanding when and how this transformation took place, as well as identifying the physical mechanisms responsible for driving it, remains one of the key open questions in extragalactic astrophysics.

One way to study this settling process in extragalactic discs is to examine their central structures, such as bars, nuclear discs, and nuclear rings. Theoretical works show that bars only form hundreds of millions of years after the dynamical settling of the disc, and that the gravitational potential of the bar can cause the gas in the interstellar medium (ISM) to shock, lose angular momentum, and funnel down to the central region. This process will be of great importance in the formation of nuclear discs and nuclear rings.

The TIMER project (Time Inference with MUSE in Extragalactic Rings; \citealt{TIMERGadotti+19}) was conceived precisely to investigate such processes by obtaining high-quality spectroscopic data for a sample of nearby barred galaxies. A key motivation is that bars can only form once discs are dynamically settled, so constraining their ages provides unique insight into the cosmic epoch of disc settling. With this approach, TIMER has delivered the first systematic bar age measurements (\citealt{desafreitas+23,desafreitas+25}), revealing both young and old bars spanning 1–13 Gyr in age. These results demonstrate that some discs were already dynamically mature at redshift $z\sim 2$, while others have only recently settled, in agreement with the observed redshift evolution of the bar fraction. Beyond bar ages, TIMER studies have also shown that bars quench star formation in their hosts and build up nuclear discs, with older bars tending to be stronger, longer, and associated with more massive central discs \citep{desafreitas+25}. TIMER has further revealed that bars are, on average, more metal-rich than their surroundings, and that the spatial distribution of stellar ages and metallicities across bars can now be reproduced in cosmological simulations (\citealt{neumann+20}).
In addition, the exquisite quality of the TIMER data enabled several breakthroughs beyond bar age properties. The project has provided the first systematic measurements of the stellar initial mass function (IMF) in star-forming regions of external galaxies, showing that while most regions are consistent with a Milky Way–like IMF, coherent spatial variations exist and can be linked to recent bursts of star formation (\citealt{martinnavarro+23}). TIMER has also uncovered evidence of stellar feedback in circumnuclear rings: in NGC~3351, young stars formed in the nuclear ring both heat the surrounding gas and drive outflows at $\sim70$ km s$^{-1}$, displacing the surrounding cold molecular gas (\citealt{leaman+19}). Moreover, the project has presented suggestive evidence that nuclear discs, rather than classical bulges, dominate the central regions of massive discs, highlighting the central role of bar-driven secular evolution in shaping galaxy centers (\citealt{gadotti+20,bittner+20}). Finally, detailed studies of nuclear bars confirm that they form and evolve in a manner analogous to main bars, strengthening the picture of a “galaxy within a galaxy,” where nuclear discs and their nuclear bars mirror the dynamics of discs and main bars at larger scales (\citealt{lorenzo+19,mendez+19,bittner+21}).

\citet{bittner+20} studied the stellar populations in nuclear discs and rings in barred spiral galaxies using TIMER data. Their results indicate that nuclear discs and nuclear rings are part of the same physical component, with nuclear rings constituting the star-forming outer edge of stellar nuclear discs. 
All nuclear discs in the sample are distinguished based on their stellar population properties, as expected in the picture of bar-driven secular evolution: nuclear discs are younger, more metal-rich, and exhibit lower [$\alpha$/Fe] enhancements compared to their immediate surroundings. The authors also find evidence of metal-poor young stellar populations in nuclear rings (or in the outer regions of nuclear discs, in galaxies in which both structures cannot be distinguished), which would be inconsistent with what is expected from the secular evolution of the interstellar medium of spiral galaxies \citep{ho+17}. However, this inconsistency has been observed in various studies that utilise spectral fitting for the young stellar populations.

Early evidence for the retrieval of young and metal-poor stellar populations can already be found in \citet{coelho+09}, where, the authors analysed high signal-to-noise integrated spectra of M32 obtained with Gemini/GMOS, using full spectral fitting techniques to disentangle its composite stellar populations. The metallicity distribution function derived spectroscopically by the authors showed a significant component with [Z/Z$_\odot$] $\leq -1$, in contrast with photometric results from resolved stars in the literature.
Although the authors do not discuss in detail the presence of young and metal-poor populations, it already anticipated the challenges that would later be more systematically addressed by subsequent studies.
\citet{carrillo+20} used pPXF and the models of \citet{vazdekis+10} to characterize the stellar populations of the galaxy NGC2903, using IFU data from the VENGA survey \citep{blanc+13}. They find that the youngest regions of the galaxy are fitted with the most metal-poor models available ([M/H] = -2.3). To alleviate this problem, they utilise a more extensive library, created by combining the \citet{vazdekis+10} templates with a younger extension from \citep{gonzalez+05}. With this altered set of templates, the authors find some contribution of models of higher metallicities in the youngest regions. \citet{emsellem+22} uses pPXF and the E-MILES \citep{vazdekis+16} to analyze the stellar populations of 19 massive star-forming galaxies from the PHANGS survey. They observe this same behaviour in the star-forming regions of these galaxies and test it using another set of models that extends to much younger ages (Charlot and Bruzual 2007, private communication; CB07). First, they exclude the younger ages of CB07 to match the same low-age limit as E-MILES. After that, they use the full CB07 set and compare the results. They show that adding models with much younger ages only marginally improves the results, and that young and low metallicity populations are still present in the stellar population maps. In \citet{pessa+23}, the authors include nebular continuum emission in the fit (which does not significantly improve the quality of the output maps), and only retrieve maps with less drastically metal-poor results by removing the lowest bins of metallicity of the sets of templates.

Recent population-synthesis studies targeting the central regions of nearby barred galaxies have provided important constraints on the nature of young and metal-poor stellar populations in nuclear environments. Using VLT/MUSE data, \citet{sextl+25b} showed that previously reported detections of extremely metal-poor young stellar populations in nuclear rings are largely methodological artifacts arising from averaging over composite stellar populations. Also, the author show that extinction maps generated through population synthesis analysis provide support for a gas infall scenario. Similar conclusions were reached in detailed population-synthesis analyses of NGC 1365 and M83 using TYPHOON spectroscopy \citep{sextl+24, sextl+25a}. In both galaxies, the young stellar component is generally super-solar in metallicity, but exhibits localized central drops that are spatially confined and interpreted as signatures of metal-poor gas infall and/or interrupted chemical evolution regulated by feedback from star formation and active galactic nuclei. Together, these studies emphasize that extreme metallicity variations inferred for young populations are highly sensitive to modelling assumptions, while supporting a physical scenario in which bar-driven inflows and feedback play a central role in shaping the chemical properties of nuclear regions.

These results highlight the need for a systematic reassessment of young stellar populations in nuclear regions using multiple stellar population models with different underlying assumptions. It is essential to discuss, however, how these models used for fitting the spectra are constructed and how they can introduce different biases to the analysis of stellar populations in galaxies. Nowadays, most models of stellar population are often built with the methodology called evolutionary synthesis \citep[e.g.][]{bc+03}. In its simplest form, it model the integrated properties of stellar populations by assembling a library of stellar spectra (theoretical or empirical), isochrones, and IMF. 

We consider as semi-empirical models the simple stellar populations (SSP) computed based on isochrones and empirical stellar spectra; conversely, we call fully theoretical models, the SSPs based on isochrones and synthetic stellar spectra \citep[see e.g. discussion in][]{coelho+20}. Differences between theoretical and empirical stellar libraries are discussed in, e.g., \citet{coelho09, coelho+07b, martins+19}. The work highlights the primary distinction between these libraries in their coverage of the Hertzsprung–Russell diagram (hereafter, HRD). Empirical libraries tend to be biased toward stars in the solar neighbourhood, whereas theoretical libraries offer the flexibility to cover a broader parameter space. Nonetheless, theoretical stellar spectra carry inherent uncertainties due to the limitations in our understanding and modelling of stellar atmospheres and opacities (see e.g. \citealt{coelho14}). 

Recent developments in stellar population synthesis have revisited this discussion in light of new comparisons between empirical and synthetic stellar libraries. \citet{coelho+20} explored the impact of the choice of stellar library on the predicted properties of galaxies, showing that the limited parameter coverage of empirical spectra has a stronger effect on predicted colours than the intrinsic imperfections of synthetic spectra. They also found that derived galaxy ages remain largely unaffected by the choice of library, while metallicities using fully theoretical models tend to be lower than those obtained from semi-empirical models. These results highlight that both approaches introduce distinct sources of bias that must be considered depending on the scientific goal.

\citet{asad+25} used the SSP models from \citet{coelho+20} to analyse a large sample of star clusters with well-constrained ages and metallicities. They found that the inferred ages and metallicities indeed depend on the adopted stellar library. Synthetic libraries tend to produce lower values for both parameters, particularly at low signal-to-noise ratios. Moreover, all models show a tendency to underestimate the ages of old, metal-poor clusters, but those based on synthetic spectra generally provide a closer match to the isochrone ages and spectroscopic metallicities available in the literature.

In this work, we aim to characterise the stellar populations in the central regions of a sample of disc galaxies studied in \citet{bittner+20}, but this time using different sets of models of stellar populations built with both theoretical and empirical stellar spectra.
In this way, we may have to consider a wider variety of possible characterisations of these stellar populations, and also study the impact of different libraries of models that may lead to these interpretations. Additionally, by measuring the gas-phase metallicity, we can compare the chemical properties of the stellar and gaseous components.

This paper is organised as follows. In Section 2, we describe the TIMER observational data used in this work. Section 3 presents our methodology, including the spectral fitting approach with pPXF, the four SSP libraries employed (MILES, X-Shooter, Starburst99, and HR-pyPopStar), and the emission-line analysis used to derive gas-phase metallicities. In Section 4, we present stellar population maps and radial profiles for our sample, organized by nuclear classification. Section 5 discusses the impact of stellar library choice on the derived stellar populations, comparing results from semi-empirical and fully theoretical models, and interpreting the young metal-poor populations in the context of gas-phase metallicities and bar-driven secular evolution. We summarize our main conclusions in Section 6.
\section{Data}

The observational data utilized in this study are part of the TIMER survey \citep{TIMERGadotti+19}. The TIMER sample comprises 21 galaxies selected from the Spitzer Survey of Stellar Structure in Galaxies ($S^4G$; \citealt{sheth+10}). The selection criteria include proximity (distance $<$ 40 Mpc), brightness ($m_B$ $<$ 15.5 mag), and angular size ($D_{25}$ $>$ 1 arcmin). All galaxies were observed with the Multi Unit Spectroscopic Explorer (MUSE) on the Very Large Telescope (VLT) in Wide Field Mode. The sample is restricted to barred galaxies exhibiting central disc-like structures as classified by \citet{buta+15}, with stellar masses exceeding $10^{10}$~M$_\odot$ and inclination angles below 60 degrees. 

We studied a subsample comprised of the galaxies NGC613, NGC1097, NGC3351, NGC4303, NGC4981, NGC4984, NGC5236, and NGC7552, all of which exhibit clear evidence of circumnuclear star-forming rings as traced by their H$\alpha$ emission maps. In addition, we include NGC1365, NGC5728, and NGC~6902, which have been classified as “peculiar” systems by \citet{bittner+20}. Table \ref{tab_galaxies_timer} provides a summary of the main properties of the galaxies analyzed in this work.

\begin{table*}
\centering
\caption{Sub-sample of TIMER galaxies studied in this work \citep{bittner+20}.}
\begin{tabular}{llcccc}
    \hline
     \textbf{Galaxy} & \textbf{Type} & \textbf{z} & \textbf{M [$10^{10} M_{\odot}$]} & \textbf{Central emission}\\
     \hline
    \\
    \textbf{Star-forming nuclear rings}\\
    \textbf{without central emission}\\
    \cline{1-1}
    NGC3351 & (R')SB(r,bl,nr)a & 0.002595 & 3.1 & SF\\
    NGC5236 & SAB(s,nr)c & 0.001733 & 10.9 & SF\\
    NGC7552 &  (R'$_1$)SB(rs,bl,nr)a & 0.005407 & 3.3 & SF\\
    \\
    \textbf{Star-forming nuclear rings}\\
    \textbf{with central emission}\\
    \cline{1-1}
    NGC0613 & SB(rs,bl,nr)b & 0.004951 & 12.2 & LINER\\
    NGC1097 & (R')SB(rs,bl,nr)ab pec & 0.004218 & 17.4 & LINER\\
    NGC4303 & SAB(rs,nl)bc & 0.005237 & 7.2 & LINER\\
    NGC4981 & SAB(s,nl)bc & 0.005602 & 2.8 & LINER\\
    NGC4984 & (R'R)SAB$_a$(l,bl,nl)0/a & 0.004116 & 4.9 & AGN\\
    \\
    \textbf{Peculiar}\\
    \textbf{nuclear regions}\\
    \cline{1-1}
    NGC1365 &(R')SB(rs,nr)bc & 0.005476 & 9.5 & AGN\\
    NGC5728 & (R$_1$)SB(r’l,bl,nr,nb)0/a & 0.009467 & 7.1 & AGN\\
    NGC6902 & (R')SAB(rs,nl)ab & 0.009363 & 6.4 & NO\\
    \hline
     \hline
     \label{tab_galaxies_timer}
\end{tabular}%
\end{table*}

\section{Methodology}

In this work, we employed the spectral fitting code \textsc{pPXF} (Penalized PiXel-Fitting, \citealt{cappellari+04,ppxf,capellari23}), which uses a maximum penalized likelihood approach to fit observed galaxy spectra as a linear combination of SSP models.

Each data cube contains approximately $10^5$ spaxels, each of which is associated with a spectrum. This enables a spatially resolved analysis of the central regions of the galaxies in our sample. However, due to spatial variations in the signal-to-noise ratio (SNR), we applied the Voronoi binning technique \citep{voronoi} to set a target SNR threshold for each spectrum used in the fits. The algorithm groups neighboring spaxels until the combined spectrum reaches a predefined minimum signal-to-noise ratio (SNR), in our case, SNR$_\text{min} = 100$, following \citet{bittner+20}.

Our analysis consists of two main steps. In the first step, we determine the stellar kinematics by fitting the Voronoi-binned spectra with \textsc{pPXF}, masking prominent emission lines to isolate the stellar continuum. This masking is performed using the \texttt{define\_goodpixels} function in \textsc{pPXF}, which excludes regions around specified emission lines from the fit. The full list of masked lines and their restframe central wavelengths is presented in Table~\ref{tab_emission_lines}. We follow the standard strategy recommended by \citet{ppxf}, which ensures a robust extraction of the stellar kinematics by avoiding contamination from nebular emission lines.\footnote{Detailed documentation available at \url{http://purl.org/cappellari/software}} The fits were run in the 4800-5800\AA\ window, following the procedure presented in \citet{bittner+20}.

The second \textsc{pPXF} run provides weights for each SSP template, from which we derive non-parametric star formation histories and calculate luminosity- and mass-weighted mean stellar population properties. We fix the kinematic parameters obtained in the first run—radial velocity, velocity dispersion, as well as the Gauss-Hermite coefficients $h_3$ and $h_4$, to prevent degeneracies, particularly between velocity dispersion and metallicity \citep[see][]{sanchez-blazquez+11}. 

An eighth-order multiplicative Legendre polynomial is included in the fitting to account for extinction effects and other broadband continuum variations. Additionally, to refine the masking in this second run, we implement a sigma-clipping routine: any wavelength region where the residuals between the observed spectrum and the first-fit model exceed three times the standard deviation (computed excluding already masked regions) is excluded from the second fit.

We used four distinct SSP libraries as template sets in our analysis, summarized below. These libraries differ not only in their spectral resolution and wavelength coverage, but also in the nature of the underlying stellar spectra. Following the distinction discussed earlier, we employed a mix of semi-empirical and fully theoretical libraries:

\begin{itemize}
    \item MILES \citep{vazdekis+15}: based on the MILES empirical stellar spectra library (\citealt{sanchez-blazquez+06}, \citealt{falcon-barroso+11}), the library provides models in 12 different metallicity values (-2.27 < [M/H] < 0.4) with ages above 0.03 Gyr, on a spectral resolution of 2.51 \AA\ (FWHM).
    
    \item X-Shooter \citep{xshooter}: based on the X-Shooter empirical stellar spectra library \citep{verro+22a}, the models span the metallicity range of -2.2 < [M/H] < +0.2 and ages above 0.05 Gyr, on a resolution of R = 10000.
    
    \item Starburst99 \citep{leitherer+99}: focused on star-forming populations, Starburst99 provides stellar population models based on theoretical stellar libraries. For more information, we refer the reader to their website\footnote{\url{https://www.stsci.edu/science/starburst99/docs/default.htm}}.
    
    \item HR-pyPopStar \citep{millan+21}: Based on three different theoretical stellar spectra libraries, the models span a wavelength range from UV to IR, with a resolution of $R_{th,5000} = 50000$ at 5000 \AA.
\end{itemize}

Additional details regarding the ingredients and parameter coverage of each SSP library are presented in Table~\ref{tab_ingredients}.

\begin{table*}
    \centering
    \caption{Emission lines fitted in this work, including individual wavelengths and, where applicable, fixed intensity ratios.}
    \label{tab_emission_lines}
    \begin{tabular}{lll}
        \hline
        Line ID & Component Lines & Wavelengths [\AA] \\
        \hline
        \texttt{[OIII]} doublet & \texttt{[OIII]}4959 / \texttt{[OIII]}5007 & 4958.92 / 5006.84 (ratio 0.33 : 1.00) \\
        HeI5876 & HeI5876 & 5875.61 \\
        \texttt{[NI]}5197 & \texttt{[NI]}5197 & 5197.90 \\
        \texttt{[NI]}5200 & \texttt{[NI]}5200 & 5200.26 \\
        \texttt{[NII]} doublet & \texttt{[NII]}6548 / \texttt{[NII]}6583 & 6548.03 / 6583.41 (ratio 0.33 : 1.00) \\
        \texttt{[OI]} doublet & \texttt{[OI]}6300 / \texttt{[OI]}6364 & 6300.30 / 6363.78 (ratio 1.00 : 0.33) \\
        \texttt{[SII]} doublet & \texttt{[SII]}6716 / \texttt{[SII]}6731 & 6716.47 / 6730.85\\
        & & (1.45 $\times$  \texttt{[SII]}6730.85 > \texttt{[SII]}6716.47 > 0.41 $\times$ \texttt{[SII]}6730.85) \\
        H$\beta$ & H$\beta$4861 & 4861.33 \\
        H$\alpha$ & H$\alpha$6562 & 6562.80 \\
        \hline
    \end{tabular}
\end{table*}

\begin{table*}
  \caption{Ingredients used for building the Simple Stellar Population models used in this work, and their parameter space range.}
  \label{tab_ingredients}

  {%
  \centering
  \begin{tabular}{llllll}
    \hline
    Library & IMF & Isochrone/Tracks & Stellar spectra & Ages [Gyr] & [M/H] \\
    \noalign{\smallskip}
    \hline
    \noalign{\smallskip}
    MILES        & K01 & BaSTI                  & MILES                  & 0.030 -- 14.0 & $-2.3 < \mathrm{[M/H]} < 0.4$ \\
    X-Shooter    & K01 & PARSEC/COLIBRI         & X-Shooter              & 0.050 -- 15.9 & $-2.2 < \mathrm{[M/H]} < 0.2$ \\
    Starburst99  & K01 & Geneva (high mass loss)& LH95, SNC02, MSH05      & 0.002 -- 14.0 & $-1.3 < \mathrm{[M/H]} < 0.4$ \\
    HR-pyPopStar & K01 & Padova                 & C14, H19, R03           & 0.001 -- 13.8 & $-0.7 < \mathrm{[M/H]} < 0.5$ \\
    \noalign{\smallskip}
    \hline
  \end{tabular}
  }

  \vspace{0.5em}
  \raggedright
  \textbf{References:}
  K01: \citealt{kroupa01};
  BaSTI: \citep{basti1,basti2};
  PARSEC/COLIBRI: \citealt{parsec}, \citealt{colibri};
  X-Shooter stellar library: \citep{xshooter_stellar};
  Geneva (high mass loss): \citealt{geneva_high};
  Padova: \citep{bressan+93,fagotto+94a,fagotto+94b,girardi+96};
  MILES stellar library: \citep{sanchez-blazquez+06};
  LH95: \citealt{leitherer+95};
  SNC02: \citealt{smith+02};
  MSH05: \citealt{martins+05};
  C14: \citealt{coelho14};
  H19: \citealt{hainich+19};
  R03: \citealt{rauch03}.
\end{table*}

For the emission-line analysis, we adopted the procedure described in \citet{silvalima+25}. In summary, the method uses the “best fit” stellar continuum obtained from the \textsc{pPXF} routine as the model for the underlying stellar population. Emission lines are then modeled with Gaussian profiles, and the observed spectrum is reconstructed by summing the stellar and emission-line components. In this work, all emission lines were fitted using the simplest configuration available in the pipeline: a single Gaussian component per line. The set of emission lines fitted is the same as in \citet{silvalima+25}.

In order to estimate the gas metallicity (here presented as the oxygen abundance), we applied the empirical S-calibration (Scal) proposed by \citet{pilyugin+16}, which relies on three standard diagnostic line ratios: N2 — $I([\mathrm{N\,II}]\lambda6548 + \lambda6584)/I(\mathrm{H}\beta)$,  
S2 — $I([\mathrm{S\,II}]\lambda6717 + \lambda6731)/I(\mathrm{H}\beta)$,  
and R3 — $I([\mathrm{O\,III}]\lambda4959 + \lambda5007)/I(\mathrm{H}\beta)$. 
For the duplets and dust extinction, we follow the same recipe as in \citet{menguiano+26} (see their section 3.4).

\section{Results}

\subsection{Stellar population maps} 
In this section, we discuss the results for the stellar population parameters in our sample. We show maps of three galaxies (NGC3351, NGC1097, and NGC1365), as representatives for each of the three groups shown in Table \ref{tab_galaxies_timer}. The remaining maps are available in the electronic appendix. 
At the average distance of the galaxies in our sample, 1" corresponds to $\approx$ 100 pc.

\subsubsection{Star forming nuclear rings without central emission}

\begin{figure*}
	\centering
	\includegraphics[width=\textwidth]{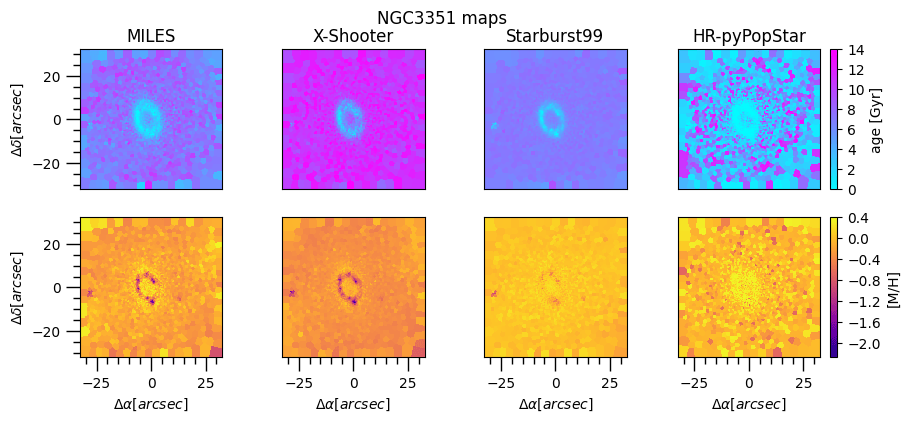}
	\caption{Maps of mean light-weighted age and [M/H] for NGC\,3351. Each column displays the maps for fits with a different library of SSP models (MILES, X-Shooter, Starburst99, and HR-pyPopStar, respectively).}
	\label{NGC3351_maps}
\end{figure*}

Figure \ref{NGC3351_maps} shows results for NGC\,3351. Maps of light-weighted mean ages are shown in the top row, and light-weighted mean metallicities in the lower row. Each column shows the best-fit results obtained from different SSP libraries, MILES, X-Shooter, Starburst99, and HR-pyPopStar, from left to right.

For NGC\,3351, the age maps clearly show the presence of a star-forming ring, characterised by ages lower than 1\,Gyr. {Three of the four libraries return smooth age gradients around the ring; HR-pyPopStar is the exception, with solutions that vary strongly from one bin to the next, giving the map a mottled appearance that reflects the instability of the fit rather than genuine small-scale structure.}
The metallicity maps show three different patterns. The two fits performed with templates based on empirical stellar spectra (MILES and X-Shooter) show the presence of a thin, irregular, and very metal-poor ring. The ring observed in the fit with Starburst99 exhibits a lower metallicity than its surroundings, but the metallicity values do not reach the extremely low values returned by the MILES and X-Shooter fits. In contrast, a more metal-rich ring can be seen in the fit performed with HR-pyPopStar. The difference in metallicity coverage and its impact on the results is explored in detail in Section~5.1.

For NGC  5236 (Figure~\ref{NGC5236_maps}), the age maps reveal a diffuse and irregular structure with a roughly circular morphology, which we identify as the nuclear disc. Nevertheless, this region is consistently younger than its surroundings, resembling the pattern observed in NGC 3351. The metallicity maps show a behavior analogous to that of NGC 3351 when semi-empirical templates are used: both MILES and X-Shooter produce irregular and very metal-poor structures, though not as clearly ring-shaped. The fits performed with Starburst99 and HR-pyPopStar also reproduce the same trends found in NGC 3351: Starburst99 indicates a metal-poor structure without reaching the extremely low values of the semi-empirical templates, while HR-pyPopStar results suggest comparatively higher metallicities, although without revealing a distinct underlying structure.

For NGC 7552 (Figure \ref{NGC7552_maps}), the age maps are similar to those of the previous galaxies, consistently showing younger central regions relative to their outskirts. The metallicity maps, in turn, closely resemble the case of NGC 5236: semi-empirical templates (MILES and X-Shooter) produce very low metallicity structures with irregular morphologies, while Starburst99 returns moderately low values, and HR-pyPopStar favors more metal-rich solutions without clear spatial structures.

Taken together, the three galaxies present broadly consistent trends. In all cases, younger stellar populations dominate the central regions, although their morphology varies from well-defined rings (NGC 3351) to irregular structures (NGC 5236 and NGC 7552).

\subsubsection{Star-forming nuclear rings with central emission}

\begin{figure*}
 	\centering
 	\includegraphics[width=\textwidth]{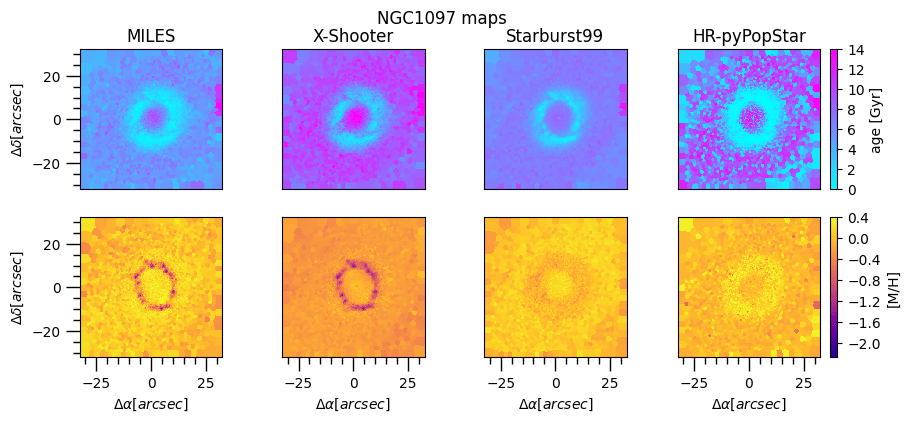}
 	\caption{Maps of mean light-weighted age and [M/H] for NGC\,1097. Each column displays the maps for fits with a different library of SSP models (MILES, X-Shooter, Starburst99, and HR-pyPopStar, respectively).}
 	\label{NGC1097_maps}
 \end{figure*}

Results for NGC 1097 are shown in Figure \ref{NGC1097_maps}, following the same layout as in Figure \ref{NGC3351_maps}. 
 
For NGC 1097, the results are analogous to those of NGC 3351, with the difference that, in this case, the metallicity maps obtained with the Starburst99 and HR-pyPopStar fits display a thicker central ring, more closely resembling the morphology seen in the age maps.

For NGC 613 (Figure \ref{NGC0613_maps}), the age maps show the same general trends described above, while the metallicity maps are also analogous for three of the libraries. The exception is HR-pyPopStar, for which the central region appears almost homogeneous, with no discernible structures. NGC 4303 (Figure \ref{NGC4303_maps}) exhibits essentially the same behavior, both in the age and metallicity maps.

NGC 4984 (Figure \ref{NGC4984_maps}) follows the same pattern in terms of stellar ages, with a young central region relative to its surroundings. However, the metallicity maps deviate from the previous cases: when fitted with MILES and X-Shooter, the central ring exhibits metallicities about 0.4 dex higher than the neighboring regions, in contrast with the consistently metal-poor solutions seen so far. For Starburst99 and HR-pyPopStar, no clear structure can be identified in the metallicity maps.

Finally, the analysis of NGC 4981 (Figure \ref{NGC4981_maps}) is inconclusive, as the low signal-to-noise ratio of the available data prevents us from drawing any robust conclusions regarding either age or metallicity.
 
\subsubsection{Peculiar nuclear regions}

\begin{figure*}
	\centering
	\includegraphics[width=\textwidth]{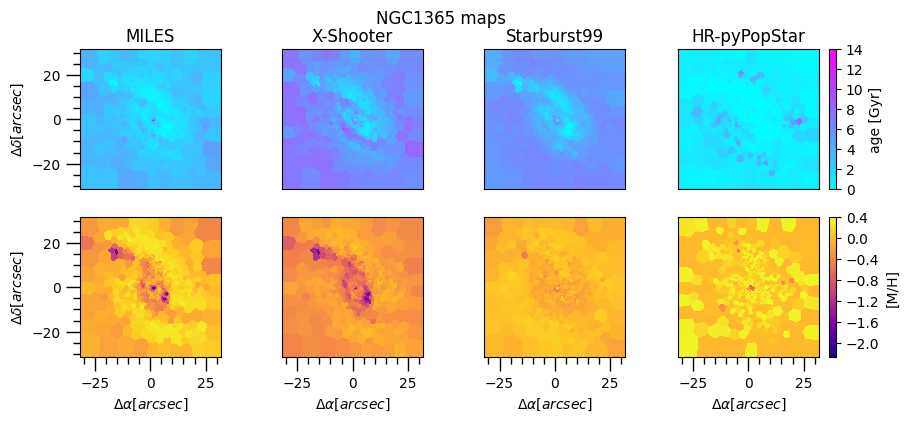}
	\caption{Mean light-weighted age and [M/H] for NGC\,1365. Each column displays the maps for fits with a different library of SSP models (MILES, X-Shooter, Starburst99, and HR-pyPopStar, respectively).}
	\label{NGC1365_maps}
\end{figure*}

Results for NGC\,1365 are shown in Figure \ref{NGC1365_maps}, following the same layout as in Figure \ref{NGC3351_maps}. 

NGC 1365 maps reveal distinct structures in the central region. The X-Shooter and Starburst99 age maps, as well as the MILES and X-Shooter metallicity maps, show the the bar dust lanes of this galaxy, with very young and metal-poor stellar populations. The age and metallicity maps from the Starburst99 fit indicate the presence of a small ring-like structure in the very center of the maps. In both maps resulting from the HR-pyPopStar fit, no clear structure is visible.

For NGC 5728 (Figure \ref{NGC5728_maps}), the age maps follow the same pattern as in the previous galaxies, with a star-forming ring composed of very young stellar populations. The surrounding regions in the HR-pyPopStar fits exhibit a mottled pattern, consistent with what is seen in other cases. The metallicity maps, however, do not reveal any discernible structures across the different templates.

NGC 6902 (Figure \ref{NGC6902_maps}) also exhibits a central star-forming ring in the age maps, with the HR-pyPopStar fits again producing mottled patterns in the surrounding regions. In the metallicity maps, the Voronoi binning allows the ring to be identified, but in three of the templates its average metallicity does not appear to differ significantly from that of the adjacent regions. The exception is HR-pyPopStar, which shows the ring as comparatively more metal-rich.

\subsection{Radial profiles}

\begin{figure*}
	\centering
	\includegraphics[width=\textwidth]{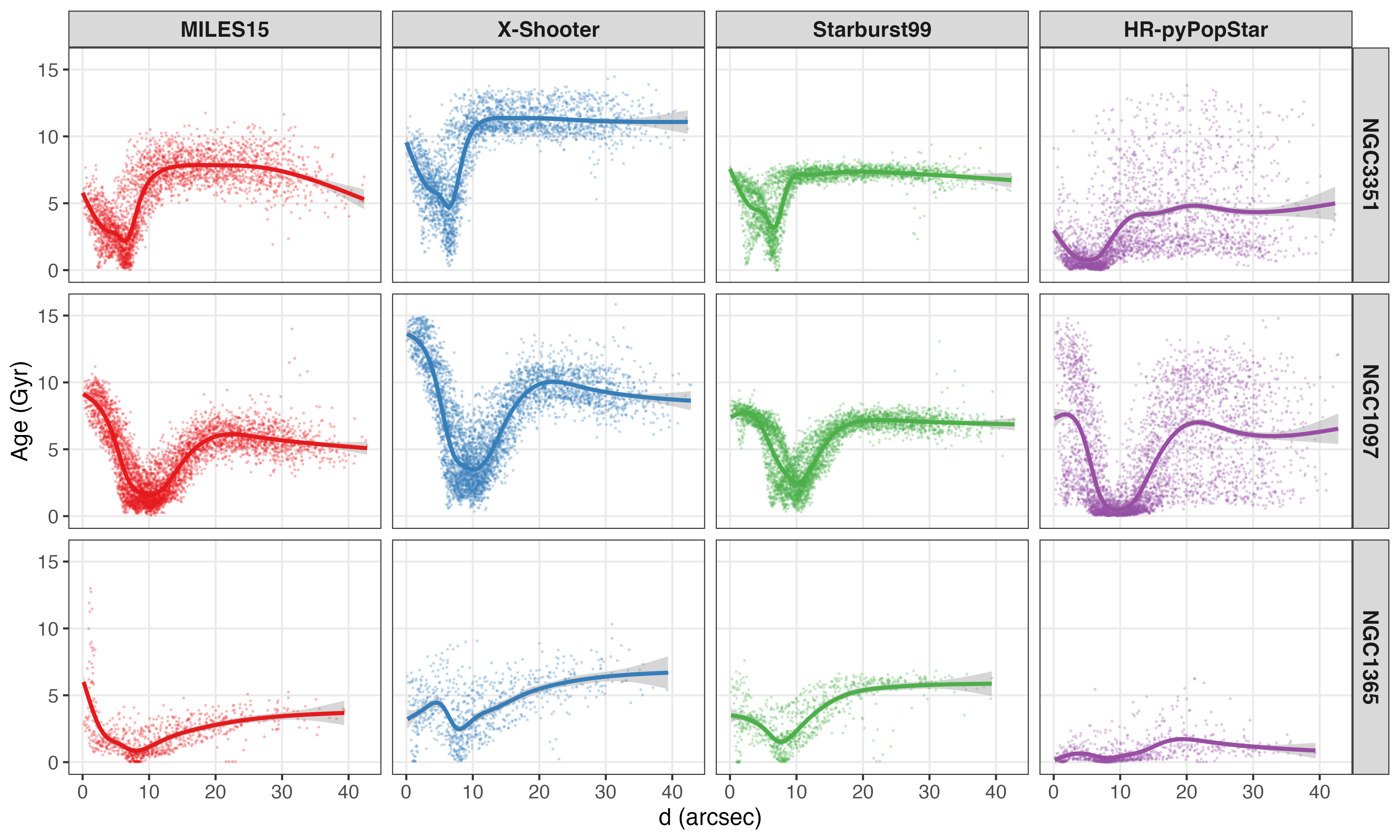}
	\caption{Light-weighted mean stellar ages as a function of distance from the galactic center for three representative galaxies: NGC\,3351 (star-forming nuclear ring without central emission), NGC\,1097 (star-forming nuclear ring with central emission), and NGC\,1365 (peculiar nuclear region). All spaxels are shown and x-axis shows the projected distance to the centre (i.e. no correction due to the inclination was applied). Each column shows results from the four SSP libraries (MILES, X-Shooter, Starburst99, and HR-pyPopStar). The profiles demonstrate that young stellar populations consistently dominate the central regions, although the absolute age values vary systematically with the choice of stellar population model. The solid lines are Locally Estimated Scatterplot Smoothing (LOESS) curves to guide the eye.}
	\label{radial_profiles_ages}
\end{figure*}

Figure \ref{radial_profiles_ages} shows the age radial profiles for the three representative galaxies in the previous section. 
The models mostly agree on the behavior of the age profile: they place the young population at the same radii and recover similar transitions to older populations in the outer disc, differing in the amplitude of age scale and scatter.
The HR-pyPopStar model exhibits the largest dispersion per radius and recovers the broadest age range, including a significant number of spaxels with ages younger than 2 Gyr. We interpret this behavior as a consequence of the more limited metallicity coverage of this library (see Table \ref{tab_ingredients}): in the absence of sufficiently metal-poor templates, the spectral fitting algorithm compensates by selecting very young stellar populations to reproduce the blue spectral features. In contrast, Starburst99 produces the narrowest dispersion and the most regular radial trends, with clear transitions from young to old populations as a function of radius. The models based on empirical libraries show intermediate scatter, with X-Shooter systematically returning the oldest ages in the central regions of NGC 3351 and NGC 1097. 

Despite the systematic offsets, all models reveal a common physical structure: for NGC 3351 and NGC 1097, a clear dip in stellar age is visible at intermediate radii ($\sim$5–15 arcsec for NGC 3351, $\sim$10–20 arcsec for NGC 1097), marking the location of the star-forming ring. This feature is detectable across all SSP libraries, providing robust evidence for recent or ongoing star formation in these annular structures. At large radii ($>$ 25–30 arcsec), all models converge toward similar ages (8–10 Gyr), indicating better agreement for older, quiescent stellar populations. In contrast, NGC 1365 shows much weaker radial structure and higher spatial scatter, consistent with its classification as a peculiar system with a more irregular star formation history.

\begin{figure*}
	\centering
	\includegraphics[width=\textwidth]{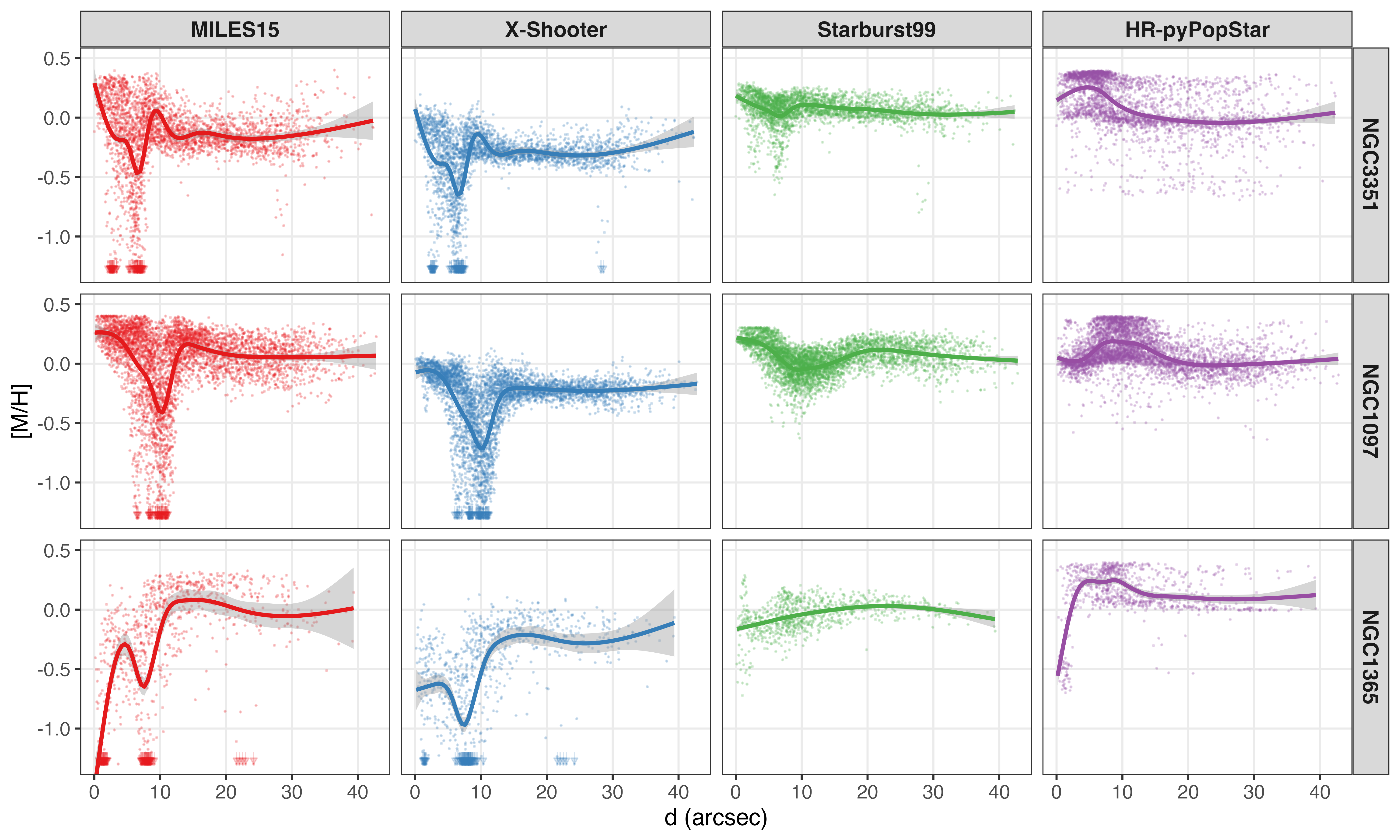}
	\caption{Light-weighted mean stellar metallicities as a function of distance to the galactic center, for the same three galaxies shown in Figure \ref{radial_profiles_ages}. 
	{The [M/H] axis is truncated at --1.3 for visual consistency across model panels; downward arrows mark spaxels whose best-fit metallicity falls below this limit.} 
	The profiles reveal systematic differences between SSP models: semi-empirical libraries (MILES and X-Shooter) systematically derive lower metallicities in the central regions, while theoretical libraries (Starburst99 and HR-pyPopStar) produce more moderate or even inverted metallicity gradients. These systematic differences highlight the strong dependence of inferred metallicity distributions on the choice of stellar population model. The solid lines are LOESS curves to guide the eye.}
	\label{radial_profiles_metallicities}
\end{figure*}

Figure \ref{radial_profiles_metallicities} presents the metallicity radial profiles, revealing larger systematic differences between models than those seen in the age profiles. The models based on empirical libraries (MILES and X-Shooter) produce sharp, V-shaped metallicity troughs in the central regions of NGC 3351 and NGC 1097, with values plunging to [M/H] $\sim$ $-1.5$ to $-2.0$ or lower—up to 2 dex below the surrounding regions. Starburst99 shows more moderate, U-shaped depressions that reach only [M/H] $\sim$ $-0.5$ to $-0.6$ at their deepest points, approximately 0.5 dex lower than the outer regions. Starburst99 also exhibits the narrowest dispersion of all models, producing the most well-defined radial trends.
The HR-pyPopStar results show a peculiar bimodal structure, with spaxels distributed into two distinct metallicity branches rather than forming a coherent radial gradient. 
This behavior likely reflects the limited metallicity coverage of this library, forcing the fit to choose between the metal-poor or metal-rich ends of the available template grid, with few intermediate solutions.

The contrast between NGC 1365 and the other two galaxies is also evident in the metallicity profiles. While NGC 3351 and NGC 1097 show well-defined central metallicity depressions across most models, NGC 1365 exhibits a more gradual and spatially diffuse metallicity decline toward the center in the MILES fit, without the sharp V-shaped feature. The HR-pyPopStar fit for NGC 1365 shows a localized metallicity enhancement (a "blob") at low radii, further highlighting the challenges this model faces in recovering coherent metallicity structures in peculiar nuclear regions. 
Beyond $\sim$25–30 arcsec, most models converge toward [M/H] $\sim$ $-0.2$ to 0.0, suggesting better agreement for the older stellar populations in the outer regions, similar to what is observed in the age profiles.

\subsection{Gas metallicities}

We applied the S-calibration on the emission line fits results in order to estimate the gas-phase metallicity in our sample. The resulting gas metallicity maps for the three proxy galaxies are presented in the upper panels of Figure \ref{gas_metal_maps_1}, with their corresponding radial profiles in the lower panels. 
Maps for the whole sample are shown in Figure \ref{gas_metal_maps}. In all maps, we show spaxels above $1\sigma$ selection threshold.

Across the three galaxies, the gas abundance is close to the solar value, with median $12+\log$(O/H) of $8.83$, $8.87$ and $8.81$ for NGC\,3351, NGC\,1097 and NGC\,1365 respectively.
For NGC 1097, the metallicity map reveals that the region where the prominent star-forming ring is located exhibits gas metallicities lower than those of the surrounding areas. 
The NGC3351 map reveals a disc-like structure with several clumpy sub-structures (see \citet{leaman+19} for more details).
NGC 1365 (described as “peculiar” by \citet{bittner+20}) presents a distinct behavior, possibly related to inner spiral arms. 

The lower panels in Figure \ref{gas_metal_maps_1} show the radial profiles. 
The gas abundance varies by $0.47$, $0.46$ and $0.61$\,dex between the 5th and 95th percentiles over $3 - 40$\,arcsec for NGC\,3351, NGC\,1097 and NGC\,1365 respectively.
Only NGC\,1097 shows a depression that is annular, with its azimuthal median reaching a minimum at $9.5$\,arcsec; the profiles of NGC\,3351 and NGC\,1365 have their minima at $4.5$ and $35.5$.

{We note that \citet{silvalima+25} compared several strong-line metallicity calibrations, including the S-calibration adopted in this work, and found that the corrected O3N2 calibration of \citet{kumari+19}, which accounts for variations in the ionization parameter, produces substantially smoother gas-phase metallicity distributions with reduced small-scale structure. Although the absolute metallicity values differ among calibrations, the corrected O3N2 maps generally exhibit more moderate spatial variations than those obtained with the S-calibration. 
This reinforces our interpretation that the large stellar metallicity contrasts inferred from the empirical SSP libraries are unlikely to reflect genuine chemical inhomogeneities in the galaxies. Instead, the gas-phase metallicity distributions, even when derived using different calibrations, consistently suggest variations of only a few tenths of a dex.

\begin{figure*}
	\centering
	\includegraphics[width=\textwidth]{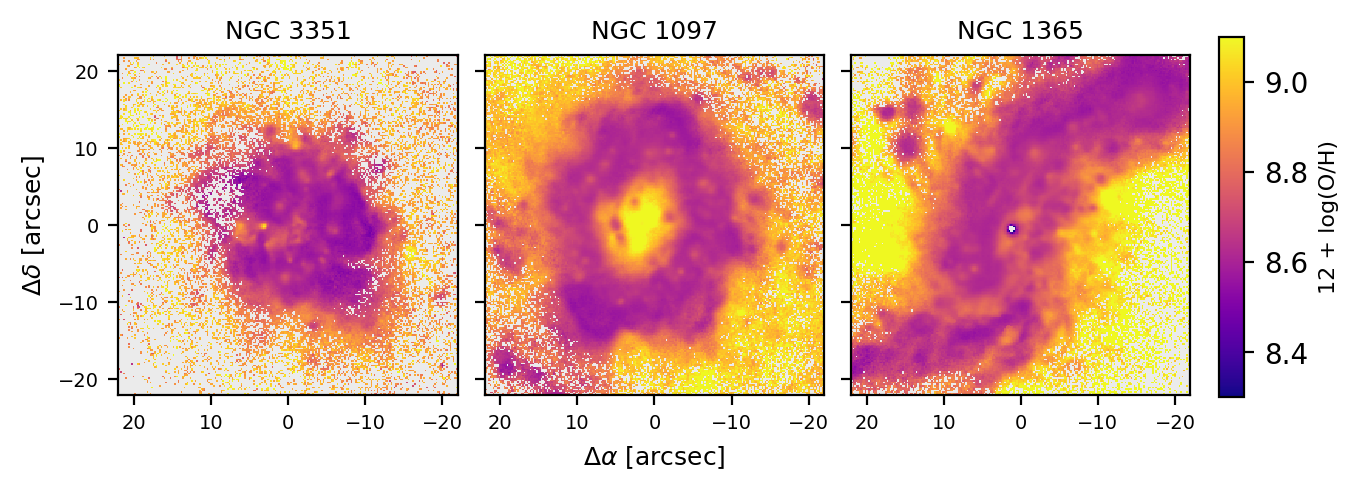}\\[2ex]
	\includegraphics[width=\textwidth]{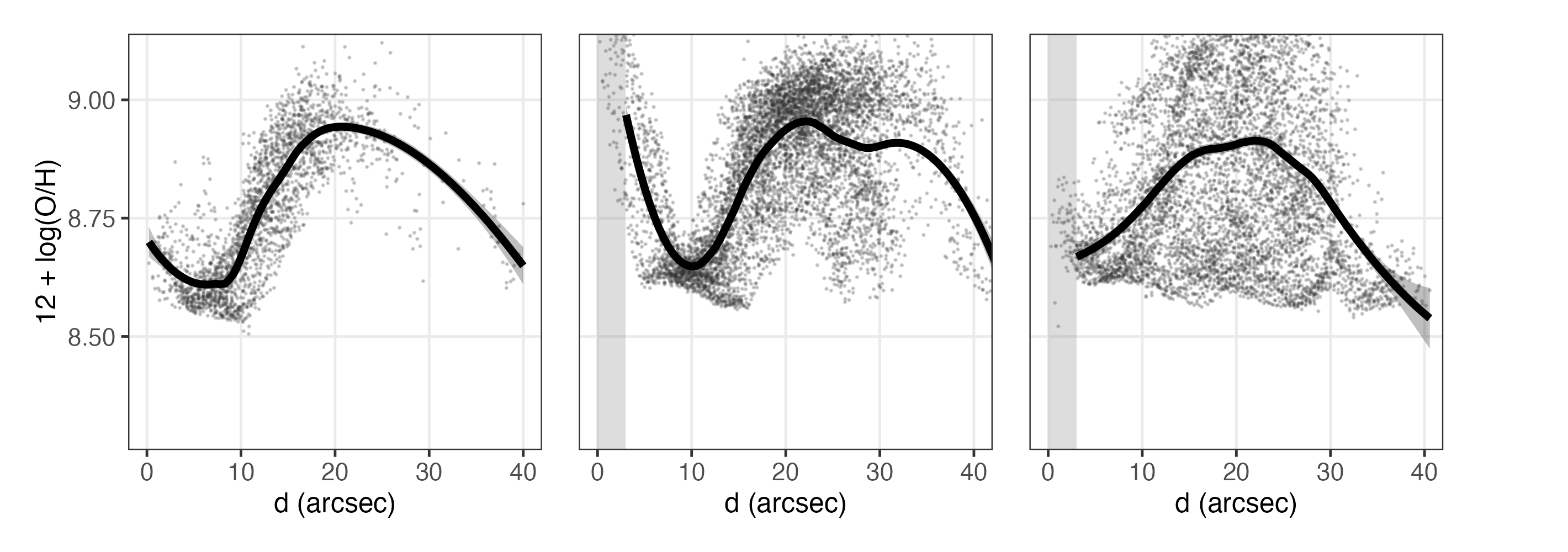}
	\caption{Gas-phase metallicity for NGC\,3351, NGC\,1097 and NGC\,1365, derived from the S-calibration. \emph{Top:} abundance maps, on a common colour scale. \emph{Bottom:} the corresponding radial profiles, showing every spaxel that survives the $1\sigma$ selection together with a LOESS smoothing and its confidence band. Radii are projected distances from the array centre, with no inclination correction. The grey band in NGC\,1097 and NGC\,1365 marks the central region affected by their Seyfert/LINER nuclei, where strong-line calibrations do not apply; points there are plotted but excluded from the LOESS fit. Between $3$ and $40$\,arcsec the abundance spans $0.47$, $0.46$ and $0.61$\,dex (5th to 95th percentile) for NGC\,3351, NGC\,1097 and NGC\,1365 respectively. }
	\label{gas_metal_maps_1}
\end{figure*}

\section{discussion}
\subsection{Parameter coverage}

One possible explanation for the discrepancies observed between the results obtained with different SSP models lies in the differences in parameter space coverage, as given in the last two columns in Table \ref{tab_ingredients}. MILES and X-Shooter libraries span a wide range of metallicities, extending down to [M/H] $\sim$ –2.3. In contrast, the Starburst99 and HR-pyPopStar libraries offer more limited coverage of metallicities. 
A similar difference is found in the age coverage, with the Starburst99 and HR-PyPopstar libraries extending to significantly younger ages than MILES and X-Shooter.

To investigate whether this limited parameter space could account for the observed differences, we constructed matched subsets of both the MILES and Starburst99 libraries, restricted to the same age and metallicity parameter space with similar parameter resolution. We define MILES (cut) and Starburst99 (cut) as subsets of their respective full libraries. 
To construct these subsets, we select the four metallicity values available in Starburst99 and the four nearest metallicity values available in MILES. We further select 53 age values between the two libraries, producing sets that span the same age parameter space and nearly identical metallicity ranges, differing by less than 0.05 dex.
The parameters of the cut sets are:

\begin{itemize}
\item Metallicities MILES: [M/H] = --1.26, --0.35, 0.06, 0.26
\item Metallicities Starburst99: [M/H] = --1.3, --0.4, 0.0, 0.3
\item Ages: \emph{t} = 0.03, 0.04, 0.05, 0.06, 0.07, 0.08, 0.09, 0.1, 0.15, 0.2, 0.25, 0.3, 0.35, 0.4, 0.45, 0.5, 0.6, 0.7, 0.8, 0.9, 1.0, 1.25, 1.5, 1.75, 2.0, 2.25, 2.5, 2.75, 3.0, 3.25, 3.5, 3.75, 4.0, 4.5, 5.0, 5.5, 6.0, 6.5, 7.0, 7.5, 8.0, 8.5, 9.0, 9.5, 10.0, 10.5, 11.0, 11.5, 12.0, 12.5, 13.0, 13.5, 14.0
\end{itemize}

Figure \ref{compara_NGC1097_maps} presents the resulting Age and [M/H] maps for NGC\,1097, derived using the full Starburst99 library (first column), our Starburst99 subset (second column), the full MILES library (third column), and the  MILES subset (fourth column).

The comparison shows that restricting the MILES library to the age and metallicity ranges of Starburst99 does not significantly alter the spatial patterns observed in the maps. While, for MILES, the metallicity values in some of the most metal-poor regions increase (as expected from the exclusion of the lowest value available in the grid) the overall structure and gradients remain largely unchanged. Importantly, the resulting maps still differ substantially from those obtained with Starburst99, indicating that the discrepancies are not solely due to differences in parameter space coverage.

In contrast, the Starburst99 metallicity maps are more strongly affected by the age–metallicity degeneracy. Restricting the available age grid alone leads to a systematic decrease in the inferred metallicities, with differences bordering 0.5 dex in some regions.

Figure~\ref{fig_cut_radial} presents the corresponding radial metallicity profiles, enabling a quantitative comparison with the profiles discussed in Section~4. The profiles reveal that the two parameter-space restrictions have qualitatively different effects on the inferred metallicity structure.
For MILES (cut), the impact is spatially localised to the star-forming ring (d\,$\sim$\,5--15\,arcsec), where the full MILES library was previously pushing to the most extreme metal-poor values. Outside the ring, the radial profile of MILES (cut) is virtually indistinguishable from the full MILES result. This localised behaviour follows directly from the nature of the metallicity restriction: it only affects solutions that were previously pinned near the edge of the available parameter space, which occurs almost exclusively in the ring.

The Starburst99 (cut) profile shows a qualitatively different response. Rather than a localised change, the age restriction produces a systematic downward shift of approximately 0.5\,dex across \textit{all} galactocentric distances, including the outer disc dominated by old stellar populations where the star-forming ring plays no role. This global character points to the age--metallicity degeneracy operating throughout the field: with fewer age templates available, the fitting algorithm compensates by adjusting metallicities across the entire population mix. The age restriction simultaneously reduces the overall density of the age grid and removes the youngest bins ($<$\,30\,Myr); the global radial extent of the shift --- spanning well beyond the ring --- suggests that grid sparseness is the dominant driver, though a contribution from the loss of very young templates cannot be excluded without additional experiments.

Crucially, even when both libraries are restricted to the same parameter space, their radial metallicity profiles remain substantially discrepant. The matched subsets do not converge toward a common solution. The experiment therefore demonstrates that the response to parameter-space constraints is itself model-dependent: for MILES the restriction is a minor perturbation localised to the ring, while for Starburst99 it introduces a new source of systematic uncertainty through the age--metallicity degeneracy. The persistent disagreement between the restricted profiles reflects fundamental differences in how the two model families encode spectral information, rather than a simple consequence of different parameter-space coverage.

\begin{figure*}
	\centering
	\includegraphics[width=\textwidth]{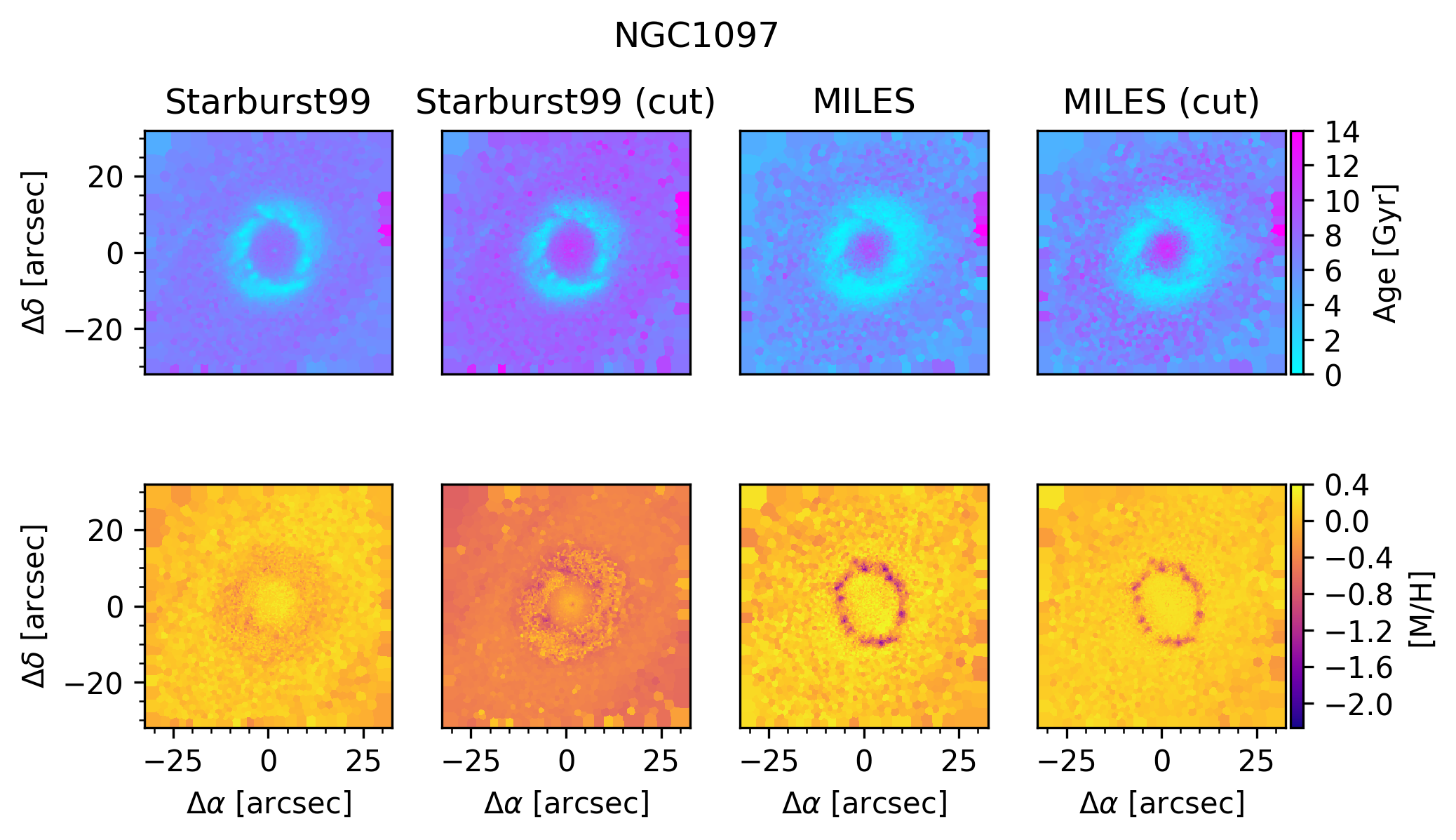}
	\caption{Comparison of stellar population maps for NGC\,1097 obtained with Starburst99 (left column), the full MILES library (middle column), and MILES with restricted metallicity coverage matching Starburst99 (right column, MILES cut). Top panels show light-weighted mean ages, bottom panels show light-weighted mean metallicities. Restricting the MILES metallicity range to match Starburst99 increases the metallicity values in the most metal-poor regions, but does not reproduce the spatial patterns obtained with Starburst99. This demonstrates that the discrepancies between semi-empirical and theoretical models are not solely due to differences in parameter space coverage.}
	\label{compara_NGC1097_maps}
\end{figure*}

\begin{figure*}
    \centering
    \includegraphics[width=\textwidth]{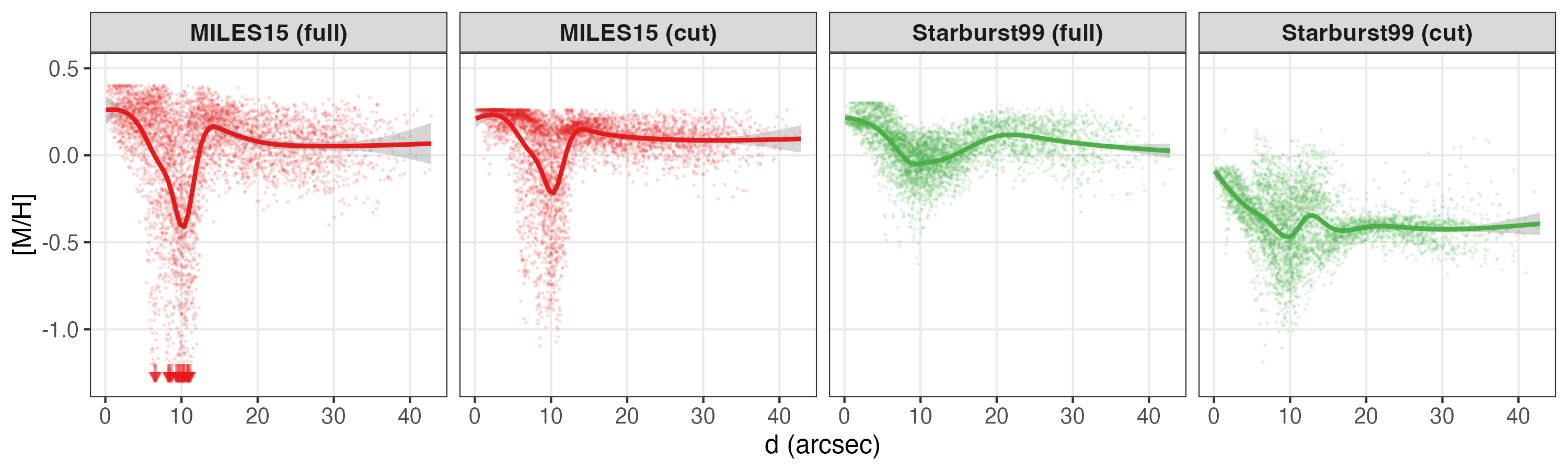}
    \caption{Radial [M/H] profiles for NGC\,1097 derived from the full and matched-subset (cut) versions of MILES and Starburst99. For MILES, the effect is localized to the star-forming ring region, with the outer profile virtually unchanged. 
    For Starburst99 the age restriction produces a global downward shift of $\sim$0.5\,dex at all radii, a consequence of the age--metallicity degeneracy operating across the entire stellar population mix. 
Downward arrows in the MILES15 (full) panel mark the spaxels with metallicity below the plotted range ($[$M/H$] < -1.3$).}
    \label{fig_cut_radial}
\end{figure*}

A complementary way to assess the role of parameter coverage is to ask how close each model's solutions come to the edge of its own metallicity grid. MILES and X-Shooter use the metal-poor quarter of grids that already extend further than the other two ([M/H] $\simeq -2.3$ and $-2.2$), confirming that their extreme solutions are not constrained by coverage. Starburst99, on the other hand, leaves two of its four metallicity nodes entirely below its recovered solutions: the moderate depressions it returns are preferred with margin to spare, not an artefact of a floor being hit. HR-pyPopStar is the opposite case, its solutions press close to the boundary of a grid spanning only $-0.7$ to $+0.4$, reinforcing that its metallicity range is too restricted for the comparison attempted here.

It is worth noting that similar tests assessing the impact of stellar libraries and parameter-space coverage have previously been performed for old stellar populations (See \citealt{martins+19}).

\subsection{Patterns across SSP models}

We aim at understanding the patterns which are introduced in the stellar populations inferences due to the choice of stellar population models. 

Figure \ref{fig-age-distributions} shows the age distributions of stellar populations across all our galaxies (each dot representing the mean age in a spaxel), revealing significant systematic differences between stellar population models. 
The HR-pyPopStar model produces a broad, relatively flat distribution extending to the youngest ages (log(age) $\sim$ 7.5). 
It is possible that the age distribution is the broadest because the metallicity range of the models is the narrowest (see Table \ref{tab_ingredients}): the age-metallicity degeneracy will force the fit to choose young populations to account for the lack of metal-poor populations among the templates.
The models based on empirical MILES show a more concentrated distribution, peaked around log(age) $\sim 9.5$, suggesting they preferentially fit intermediate-age stellar populations. 
Starburst99 exhibits a sharp, narrow peak at log(age) $\sim 9.8$, while X-Shooter peaks at similar values and extends towards older ages. 
These systematic variations highlight the significant model-dependent uncertainties in stellar population age determinations.

\begin{figure}
	\includegraphics[width=\columnwidth]{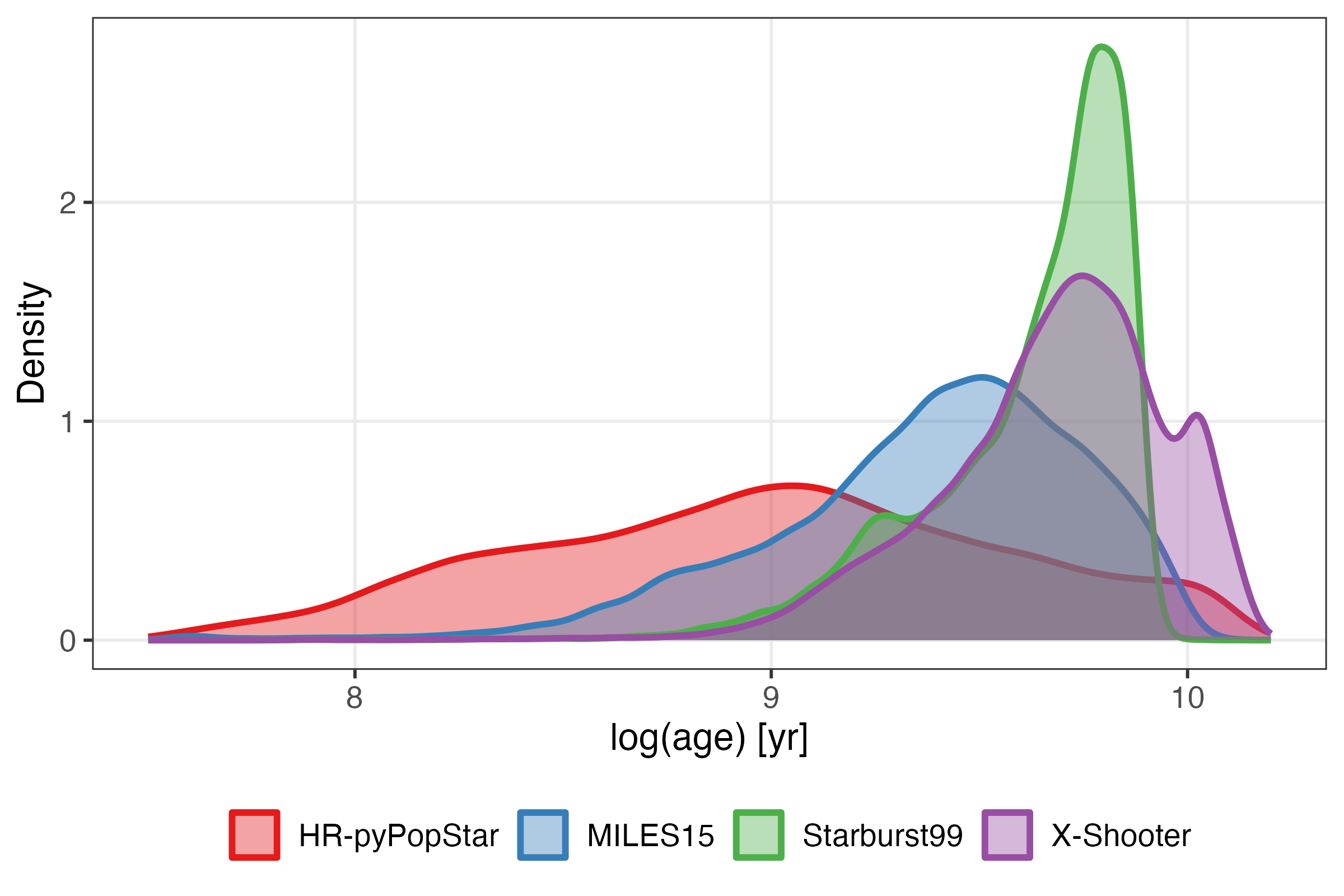}
	\caption{Age distributions of stellar populations derived from spectral fitting across all 11 TIMER galaxies, for each stellar population model. The distributions show systematic differences between models, with the HR-pyPopStar high-resolution theoretical library recovering the broadest age range, including very young populations, the semi-empirical MILES library favouring intermediate ages, and the Starburst99 evolutionary synthesis model and X-Shooter favouring older stellar populations. Only spaxels with log(age) > 7.5 are included.}
	\label{fig-age-distributions}
\end{figure}

Figure \ref{fig-metallicity-distributions} displays similar visualizations for the metallicity distributions.
The HR-pyPopStar results show a distinctly bimodal distribution with peaks around [M/H] $\sim$ -0.3 and [M/H] $\sim$ +0.2, not seen with the other models. 
This behaviour might be due to the limited range of metallicities.
The Starburst99 model exhibits a sharp, narrow peak centered near solar metallicity ([M/H] $\sim$ 0.0), indicating a possible bias toward solar-abundance stellar populations. 
The other libraries show intermediate behaviour: X-Shooter displays a broad distribution extending to more metal-poor values with a peak around [M/H] $\sim$ -0.4, while MILES shows a more concentrated distribution peaked near [M/H] $\sim$ -0.2.

\begin{figure}
	\includegraphics[width=\columnwidth]{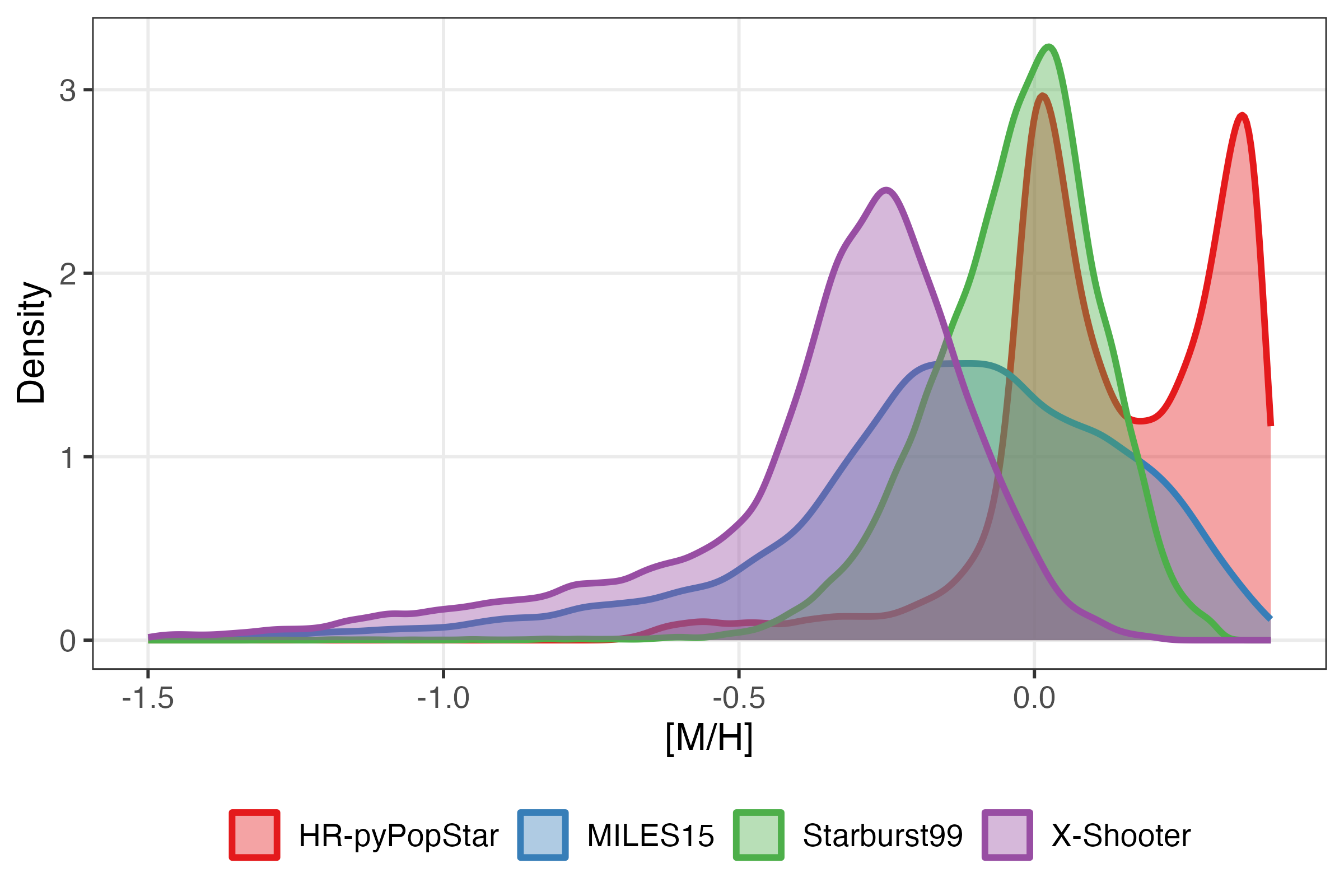}
	\caption{Metallicity distributions of stellar populations derived for the whole sample, for each stellar population model. The distributions reveal systematic model differences: the HR-pyPopStar high-resolution theoretical library shows a bimodal distribution, Starburst99 is strongly peaked near solar metallicity, X-Shooter shows the lowest metallicity peak, and MILES shows a broad metallicity distribution.}
	\label{fig-metallicity-distributions}
\end{figure}

\subsection{Stellar populations for different nuclear types}

\begin{figure}
	\includegraphics[width=\columnwidth]{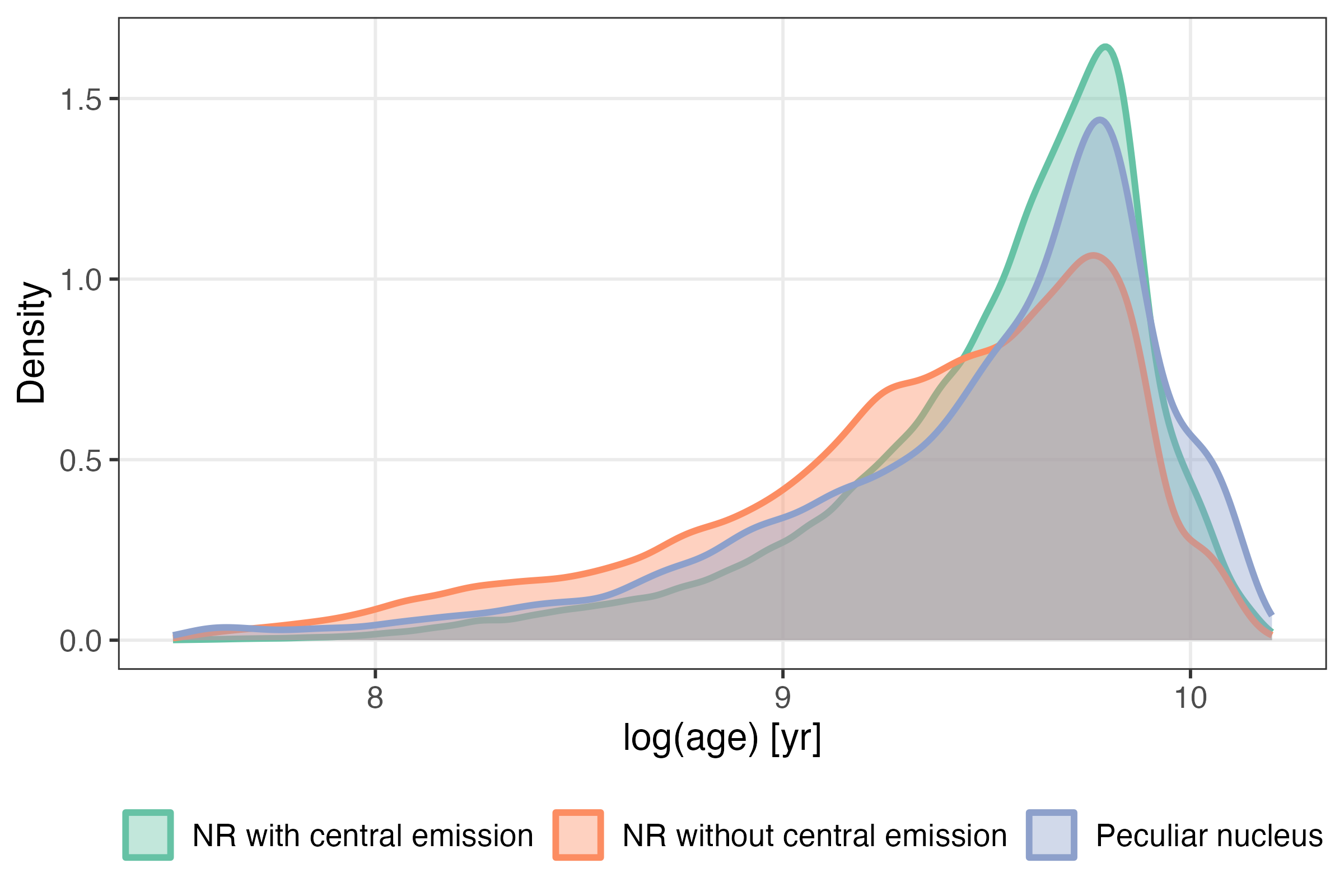}
	\caption{Age distributions across all galaxies grouped by nuclear classification. All three categories show similar peaked distributions around log(age) $\sim$ 9.8, yet nuclear rings without central emission show an excess of younger populations. Only spaxels with log(age) > 7.5 are included.}
	\label{fig-ages-by-category}
\end{figure}

Figure \ref{fig-ages-by-category} shows the age distributions of stellar populations across the galaxy sample, grouped by nuclear classification (nuclear rings without central emission, nuclear rings with central emission and peculiar nuclear regions). 
The three galaxy categories display remarkably similar age distribution patterns, all peaking around log(age) $\sim 9.8$ (corresponding to $\sim$6 Gyr stellar ages). 
Nuclear rings without central emission (orange) exhibit an excess tail extending to younger ages (log(age) $\sim 8 - 8.5$ ) compared to the distribution of nuclear rings with central emission (green). This suggests that these systems experienced more recent star formation episodes, or, equivalently, the central emission correlates with a faster quenching \citep[see e.g. discussions in][]{coelho+11}.
Results from the four stellar population models are shown in the figure, with the intention of highlighting patterns that are robust against the choice of models. Indeed, we verified that when plotting results from the SSP models individually, the pattern is the same.

We verified with statistical tests that the differences are significant (Kruskal-Wallis p < 2.2e-16; all pairwise comparisons p < 0.003), albeit with effect sizes being small (Cohen's d values range from 0.125 for the peculiar nuclei vs. nuclear rings with central emission to 0.433 for the nuclear rings without vs. with central emission). The biggest difference appears between the two nuclear ring categories: galaxies without central emission show younger mean ages compared to those with central emission, corresponding to a small but measurable effect (d = -0.433). The highly significant p-values, despite small effect sizes, reflect the large sample size (n > 100,000 spaxels). While real differences exist, they are subtle.

The same analysis for metallicities, shown in Figure \ref{fig-metallicities-by-category}, shows a more nuanced behaviour. The figure displays the metallicity distributions coloured by nuclear classification. 
All three categories show broad distributions centered around slightly sub-solar metallicities ([Fe/H] $\sim$ -0.2), but with distinct characteristics.

Contrary to the case of age distributions, some of the features seen in this visualization were identified to be model-dependent. This finding implies that model-dependent metallicity uncertainties dominate over real astrophysical variations between nuclear ring classifications in the TIMER sample.	
The specific metallicity ranges and peak positions vary between models. The HR-pyPopStar model shows a distinctive bimodal structure displaying two clear peaks (around [Fe/H] ~ -0.2 and +0.2). However, this bimodal pattern largely disappears with the other models. The Starburst99 model produces the most similar distributions across categories.
While the overall metallicity scales and peak values are model-dependent, the relative breadth of metallicity distributions across categories appears to be a robust astrophysical signature. Nuclear rings without central emission consistently show a tail towards low metallicities in all SSP models but HR-pyPopStar, supporting the interpretation that nuclear classification do reflect real differences in chemical evolution history, but with systematic uncertainties in absolute metallicity determinations dominating the observed distributions.

\begin{figure}
	\includegraphics[width=\columnwidth]{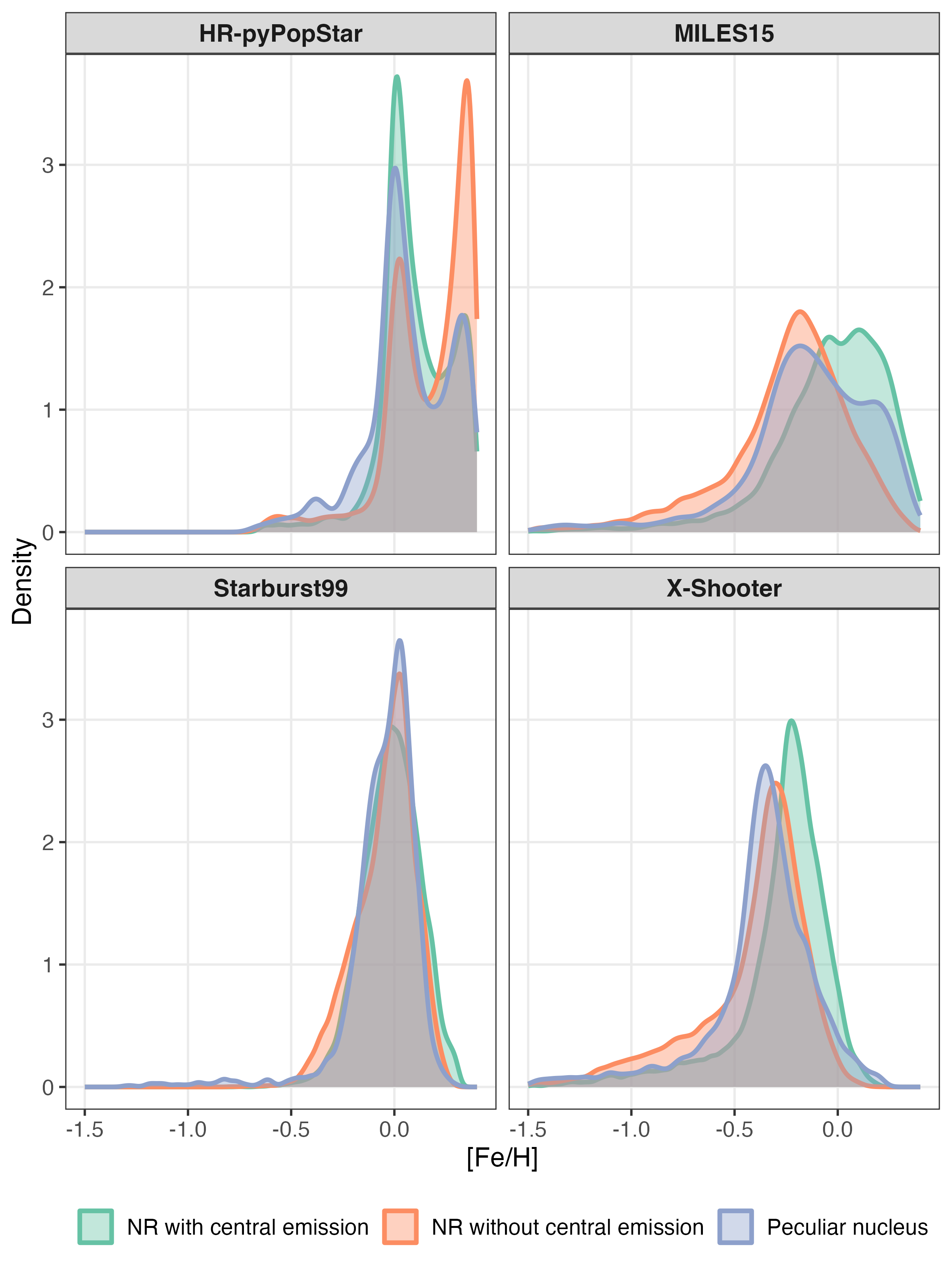}
	\caption{Metallicity distributions across all galaxies grouped by nuclear classification, shown separately for each stellar population model. The distributions reveal model-dependent systematics in the retrieved metallicity values. HR-pyPopStar shows a distinctive bimodal structure with peaks around [Fe/H] $\sim$ -0.2 and +0.2. MILES produces a broader distribution extending to lower metallicities. Starburst99 yields the most similar distributions across all three nuclear classifications, with narrow peaks around solar metallicity. X-Shooter shows intermediate behavior. This comparison demonstrates that differences between nuclear classifications are smaller than systematic differences between models.}
	\label{fig-metallicities-by-category}
\end{figure}

\subsection{The young and metal-poor populations in TIMER galaxies}

Previous studies of stellar populations in galaxies from the TIMER sample have reported the presence of very young components with extremely low metallicities, up to 2 dex lower than their immediate surroundings. These results were primarily obtained using semi-empirical stellar population models, which are known to suffer from limited coverage in certain regions of the Hertzsprung-Russell diagram. One of the most significant gaps lies in the representation of very young, massive stars, which can lead to strong biases in the inferred properties of young stellar populations.

In our analysis, we confirm that the MILES and X-Shooter models reproduce these
earlier findings, revealing young stellar components with metallicities much
lower than those of nearby regions. 
Starburst99 also recovers young
populations more metal-poor than their surroundings, but the difference is more
moderate, on the order of a few tenths of a dex. HR-pyPopStar inverts the
trend, returning young populations slightly more metal-rich than the older ones.

To provide an independent reference, we also derived gas-phase metallicities.
Before comparing the stellar results with the gas-phase metallicities, an important caveat is in order. 
The two abundances measure different quantities: 
$12+\log$(O/H) is an oxygen abundance on an absolute scale, whereas [M/H] is a total metallicity
defined by each library's own solar reference, and the empirical libraries
carry the Milky Way [$\alpha$/Fe]--[Fe/H] relation of their stars while the
Geneva and Padova tracks behind Starburst99 and HR-pyPopStar are scaled-solar.
The offset of about $0.2$\,dex between the MILES and X-Shooter results in Figure \ref{radial_profiles_metallicities}
is itself a measure of different zero points. 
A further difference is that the gas traces the present-day
interstellar medium, while the stellar metallicity is light-weighted over the
whole star formation history sampled by each bin; the two need not coincide
even for a perfect model. 
These difficulties all concern the metallicity scale,
and they can be set aside by comparing the \emph{geometry} of the radial
features rather than their amplitude.

We make this comparison for NGC\,1097, the one galaxy of the three in
which the depression is unambiguously annular. The gas depression is centred at
$9.5$\,arcsec and is about $10$\,arcsec across at half its depth (Figure
\ref{radial_profiles_metallicities}). Starburst99 places a depression at the
same radius and reproduces its width to within an arcsecond. MILES and
X-Shooter also find the correct radius, at $10.5$\,arcsec, but compress the
feature into $3$ and $5$\,arcsec 
(between a third and a half of the width of the gas feature, see Figure \ref{gas_metal_maps_1}),
and they overshoot to metal-rich values at
$13$--$17$\,arcsec where the gas is still depleted. HR-pyPopStar returns a
feature at $9.5$\,arcsec with approximately the correct width, but of the
opposite sign, recovering the ring as metal-rich where the gas is metal-poor.
Of the
four libraries, only Starburst99 reproduces the radius, width and sign of the
gas depression simultaneously. MILES and X-Shooter find the correct radius but
compress the feature to a third of its width, and HR-pyPopStar matches the
radius and width but inverts the sign.

Figure~\ref{fig-stellar-gas-proxies} shows the relationship between
stellar and gas-phase metallicities for the three representative galaxies.
Over these three galaxies the gas spans $0.49$\,dex between the 5th and 95th
percentiles, against $1.11$ and $0.98$\,dex for MILES and X-Shooter and $0.40$
and $0.49$\,dex for Starburst99 and HR-pyPopStar. The MILES and X-Shooter
panels are L-shaped: one branch along which the gas abundance varies at nearly
constant stellar metallicity, and a second in which [M/H] changes by more than
a dex while the gas abundance barely moves. It is the second branch that
produces the extreme values, and it has no counterpart in the gas. The
Starburst99 and HR-pyPopStar panels instead form compact clouds with little
stellar variation of any kind.

\begin{figure*}
	\includegraphics[width=\textwidth]{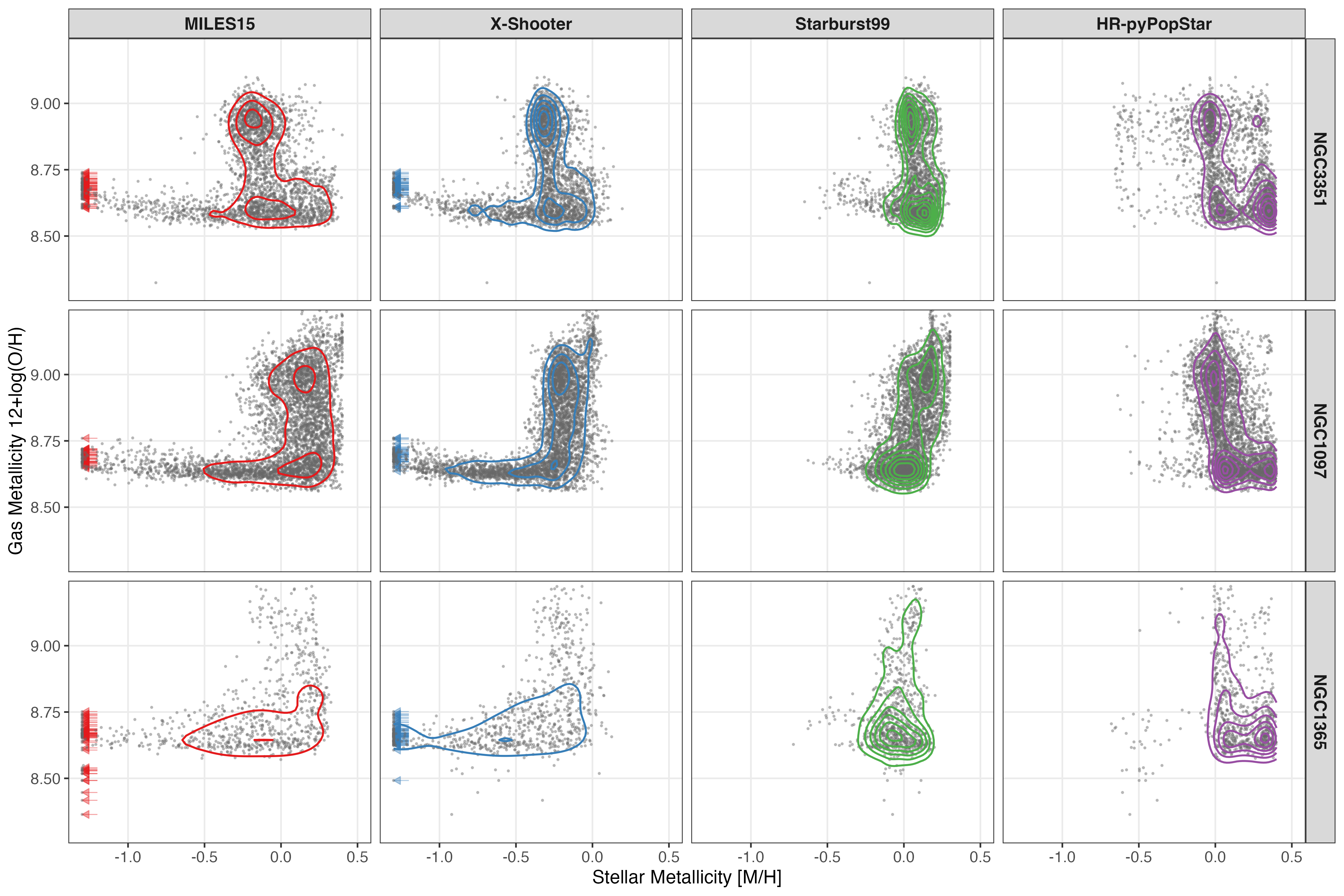}
	\caption{Stellar metallicity against gas-phase metallicity for the three representative galaxies, by SSP model (columns) and galaxy (rows). Each point is one Voronoi bin; the gas value is the median of the spaxels within it, from the S-calibration with the $1\sigma$ selection, and contours show the density of points. Arrows mark bins whose stellar metallicity falls below the plotted range.}
	\label{fig-stellar-gas-proxies}
\end{figure*}

It is worth returning at this point to the spatial distribution of the extreme values. In the metallicity maps of NGC\,1097 (Figure \ref{NGC1097_maps}) the most metal-poor solutions returned by MILES and X-Shooter are not distributed smoothly around the star-forming ring but appear as a series of discrete clumps along it, while the ring recovered by Starburst99 and HR-pyPopStar is smooth and shallow. The extreme metallicities are therefore a localised substructure found only in the fits with empirical libraries, rather than a property of the ring as a whole.
A spatial correspondence between these extreme metal-poor
solutions and the presence of Wolf-Rayet features has been noted within the
collaboration (Boecker et al., in preparation). Should this be confirmed, it
would offer a natural explanation for the behaviour we report: the empirical
libraries adopted here are built from Milky Way field stars and contain no
Wolf-Rayet atmospheres, whereas both theoretical libraries include them by
construction (see Table~2).

Our results are also in line with recent population-synthesis studies of
nuclear star-forming regions in barred galaxies. Using VLT/MUSE data,
\citet{sextl+25b} showed that previously reported detections of extremely
metal-poor young stellar populations in nuclear rings are largely
methodological artefacts arising from the averaging over composite stellar
populations. By disentangling the young and old components, they found that the
young stellar population is typically near-solar or super-solar in metallicity,
while exhibiting localised and moderate decreases in metallicity that are
spatially confined and interpreted as signatures of metal-poor gas inflow.
Similar conclusions were reached in detailed population-synthesis analyses of
NGC~1365 and M83 based on TYPHOON spectroscopy
\citep{sextl+24,sextl+25a}. {Their finding that composite-population
averaging is itself a source of spurious metal-poor solutions bears directly on
the comparison in Figure~\ref{fig-stellar-gas-proxies}, since our stellar
metallicities are light-weighted over the full population mix in each bin. It
supports the interpretation that the extreme metallicity ranges inferred from
the semi-empirical models are predominantly driven by modelling systematics
rather than by the chemical properties of the young stellar populations.}

The presence of young, metal-poor stellar populations in the central
regions of TIMER galaxies is usually explained by internal gas flows driven by the bar: the inflow increases the concentration of molecular gas in the central
parts and enhances star formation, and the newly formed stars inherit the lower
metallicity of the inflowing gas \citep{menguiano+26}. 
Our measurements show gas abundances depression of a few tenths ($\sim 0.3$\,dex in the ring of NGC\,1097), while abundance gradients in spiral discs are typically $-0.04$ to $-0.07$\,dex\,kpc$^{-1}$
\citep{vanzee+98, sanchez+14, pilyugin+14, sanchez+16, belfiore+17, sanchez+18}.
The amplitude of the gas depression therefore exceeds what bar-driven
redistribution of disc material alone can supply, and another (currently unknown) source of gas is needed to explain our findings.

\subsection{Stellar Age-Metallicity relations}

Figure~\ref{fig-amr-proxies} shows the age--metallicity relations for
the three representative galaxies. All four models recover a two-component
structure. The first component follows the canonical self-enrichment pattern, in
which younger populations are progressively more metal-rich as successive
generations of stars return processed material to the interstellar medium. The
second component runs in the opposite direction, with younger populations more
metal-poor than older ones, and its presence requires that the star-forming gas
was diluted or replaced by lower-metallicity material at some point in the
recent past. It is the shape and amplitude of this second component that varies
most strongly between models.

The semi-empirical libraries produce a steep, monotonic decline in the
young branch with a distinctive ``hook'' shape, particularly pronounced in
NGC\,3351 and NGC\,1097.
MILES and X-Shooter results decline by  $> 1$ dex from the oldest to the youngest populations. 
Starburst99 also recovers a metal-poor young component, $\sim 0.1 -- 0.2$\,dex more metal-poor than those of intermediate age.
HR-pyPopStar shows both components following in-situ-like enrichment patterns, compressed into a narrow metallicity range.

The different models call for gas infall scenarios of different kinds. 
The pattern recovered by the semi-empirical models requires the supply of metal-poor gas to be sustained over time, outpacing its own chemical enrichment throughout the episode, delivering ever more pristine material as star formation proceeds. The pattern recovered by Starburst99 is consistent with a single dilution episode: gas arrives, the metallicity of the star-forming material is lowered once, and enrichment then proceeds normally.
Independent evidence disfavours the first scenario:
\citet{menguiano+26} find that the circumnuclear H\,{\sc ii} regions of the
TIMER galaxies have \emph{higher} N/O than those of the discs, which points to
a declining rather than a rising star formation rate in these regions. 
A large, sustained supply of pristine gas is difficult to reconcile with that signature. 
Together with the amplitude of the gas depression measured in Section 4,
which is a few tenths of a dex, this argues that the steep monotonic branch recovered by the semi-empirical libraries is not tracing a physical dilution history. The metallicity variations recovered by Starburst99 are of the same
order as those of the gas, and the shape of its young branch is
that of a single dilution episode rather than a sustained one. Nei-
ther observation establishes that its age–metallicity relations are
correct, but both leave them compatible with the gas in a way
the semi-empirical relations are not.

\begin{figure*}
	\includegraphics[width=\textwidth]{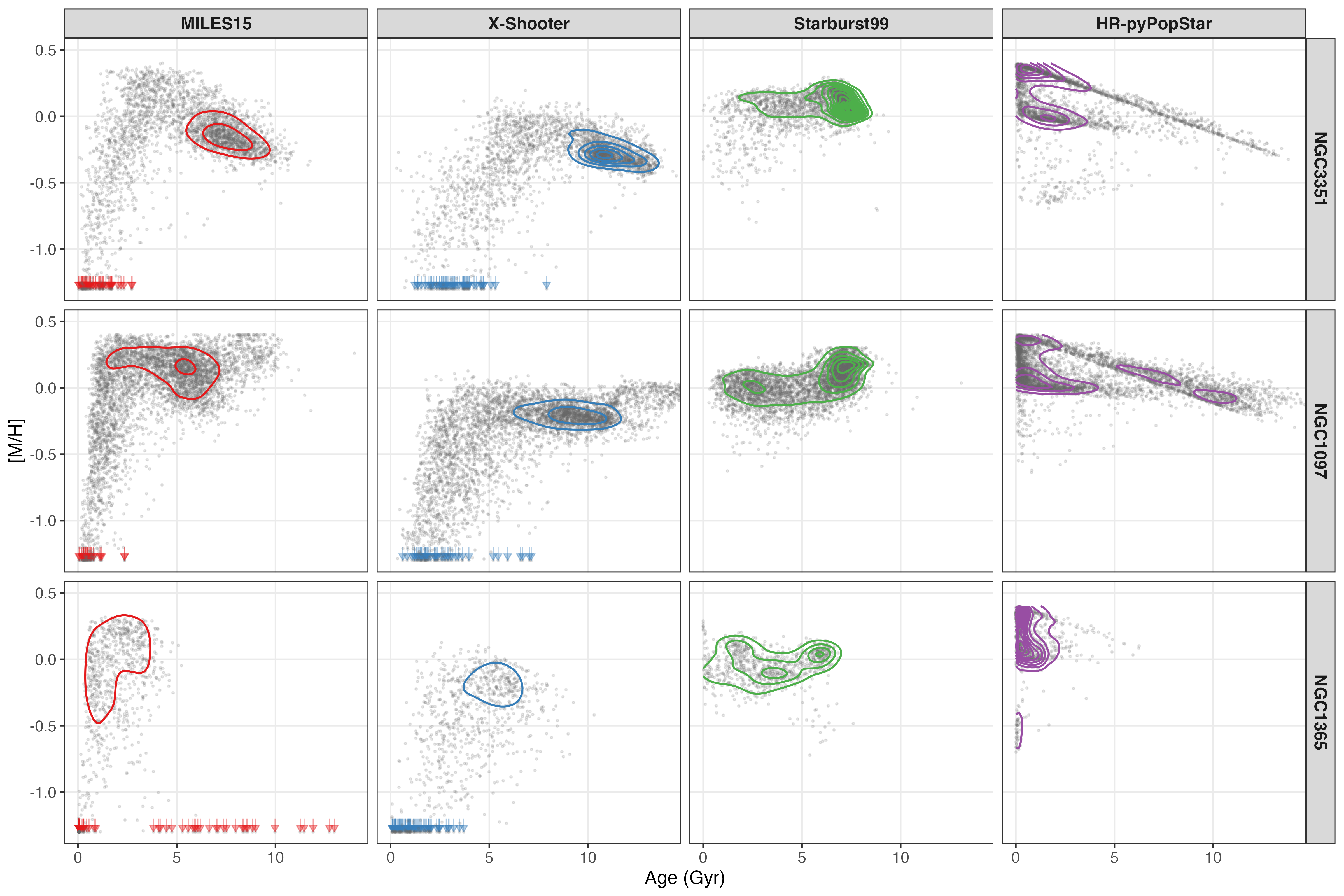}
	\caption{{Age--metallicity relations for NGC\,3351, NGC\,1097 and NGC\,1365. Each row corresponds to one galaxy and each column to one SSP library. The semi-empirical libraries (MILES, X-Shooter) produce a steep ``hook'' in which younger populations are progressively more metal-poor; Starburst99 shows a departure only in its youngest bin; HR-pyPopStar returns younger populations that are more metal-rich than older ones, with no hook.}}
	\label{fig-amr-proxies}
\end{figure*}

\section{Conclusions}

In this work, we have analyzed the stellar populations of a subsample of TIMER galaxies using four different stellar population models, built on empirical (MILES, X-Shooter) and theoretical (Starburst99, HR-pyPopStar) stellar flux libraries. Our goal was to evaluate how the choice of model affects the inferred properties of young stellar populations in the central regions of disc galaxies. We find that all galaxies display young stellar components concentrated in their central regions (either as well-defined rings or irregular structures).

A key result of this study is the strong model dependence of the derived metallicities. Fits based on semi-empirical templates yield young populations far more metal-poor than their surroundings, whereas the theoretical libraries return more moderate values. Measured consistently as the 5th-to-95th percentile spread of the recovered metallicities across the three galaxies analysed in detail, MILES and X-Shooter vary by $1.11$ and $0.98$\,dex against the $0.49$\,dex spanned by the gas in the same regions, while Starburst99 and HR-pyPopStar vary by $0.40$ and $0.49$\,dex. The integrated-light stellar metallicity is an average over different stellar populations, though the observations remain dominated by the youngest ones, while the gas metallicity measures the present-day interstellar medium from which stars are forming; the two are therefore expected to be related, but they are not the same quantity, and their zero points are set differently by each library. The comparison excludes the largest stellar variations returned by the semi-empirical libraries, which exceed anything present in the gas, without thereby validating the absolute scale of any model.

The age--metallicity relations reveal a two-component structure across most galaxies and models, reflecting the complex star formation history in these barred systems. The two families of models imply qualitatively different scenarios: the semi-empirical libraries recover a monotonic decline of $> 1.0$\,dex from old to young populations, which would require a sustained inflow of metal-poor gas outpacing its own enrichment, whereas Starburst99 departs from the older populations only at the most recent ages and by a few tenths of a dex, consistent with a single dilution episode. The gas-phase abundances and the nitrogen-to-oxygen ratios measured by \citet{menguiano+26} disfavour the sustained scenario, arguing that the steep branch recovered by the semi-empirical libraries is not tracing a physical enrichment history. The age--metallicity relations from the theoretical libraries remain compatible with the gas measurements in a way the semi-empirical ones are not.

Overall, our results highlight that stellar population inferences—particularly age-metallicity relations—are subject to significant systematic uncertainties driven by the adopted SSP model. The combination of semi-empirical and theoretical templates, together with independent gas-phase metallicity constraints, offers a more robust framework for disentangling the physical properties of young stellar populations. 

\section*{Acknowledgements}

GG is supported by Coordenação de Aperfeiçoamento de Pessoal de Nível Superior - Brasil (CAPES) - Finance Code 88887.509924/2020-00 and Conselho Nacional de Desenvolvimento Científico e Tecnológico - Brasil (CNPq) - Finance Code: 200958/2022-4.\\
PRTC acknowledges the support by Conselho Nacional de Desenvolvimento Científico e Tecnológico (Process 308715/2025-0). \emph{AI transparency:} PRTC collaborated with Claude AI (Anthropic) throughout this project on data analysis, code implementation, and iterative manuscript writing.\\
LASL is supported by Coordenação de Aperfeiçoamento de Pessoal de Nível Superior - Brasil (CAPES) - Finance Code 88887.637633/2021-0 and by CNPq grants 400990/2022-9, 315206/2026-8, and 402281/2026-8.\\
LPM thanks the Fundação de Amparo à Pesquisa do Estado de São Paulo (FAPESP, grant 2025/06495-9) and CNPq (grant 307056/2025-2) for partial funding of this research.\\
MM acknowledges support from the UK Science and Technology Facilities Council through grant ST/Y002490/1.\\
D.A.G. is supported by the UK Science and Technology Facilities Council (STFC) grant ST/X001075/1.\\
JMA acknowledges the support of the Agencia Estatal de Investigación del Ministerio de Ciencia e Innovación (MCIN/AEI/10.13039/501100011033) under grant nos. PID2021-128131NB-I00 and CNS2022-135482 and the European Regional Development Fund (ERDF) ‘A way of making Europe’ and the ‘NextGenerationEU/PRTR’.\\
AdLC acknowledges financial support from the Spanish Ministry of Science and Innovation (MICINN) through RYC2022-035838-I and PID2021-128131NB-I00 (CoBEARD project). 

\section*{DATA AVAILABILITY}

The raw and reduced MUSE data used in this work are publicly available at the ESO Science Archive Facility.

\bibliographystyle{mnras}
\bibliography{references} %

\clearpage
\onecolumn

\appendix

\section{Appendix}
\subsection{Stellar population maps for the whole sample}

Figures \ref{NGC5236_maps} to \ref{NGC6902_maps} show the stellar population maps for the galaxies not shown in the main text.

\begin{figure*}
	\centering
	\includegraphics[width=\columnwidth]{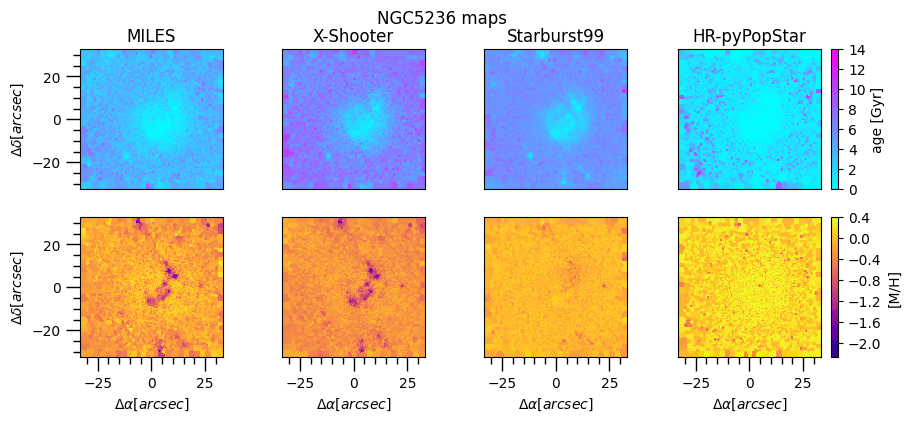}
	\caption{Same as Figure \ref{NGC3351_maps}, but for NGC\,5236.}
	\label{NGC5236_maps}
\end{figure*}

\begin{figure*}
	\centering
	\includegraphics[width=\columnwidth]{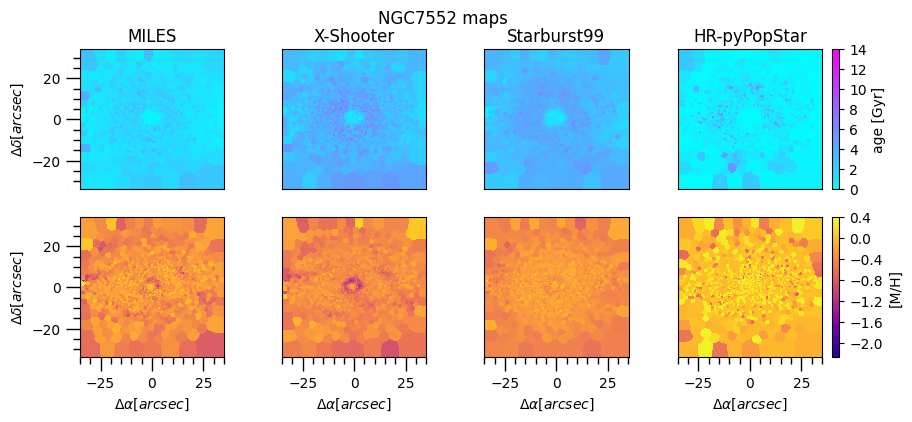}
	\caption{Same as Figure \ref{NGC3351_maps}, but for NGC\,7552.}
	\label{NGC7552_maps}
\end{figure*}

\begin{figure*}
	\centering
	\includegraphics[width=\columnwidth]{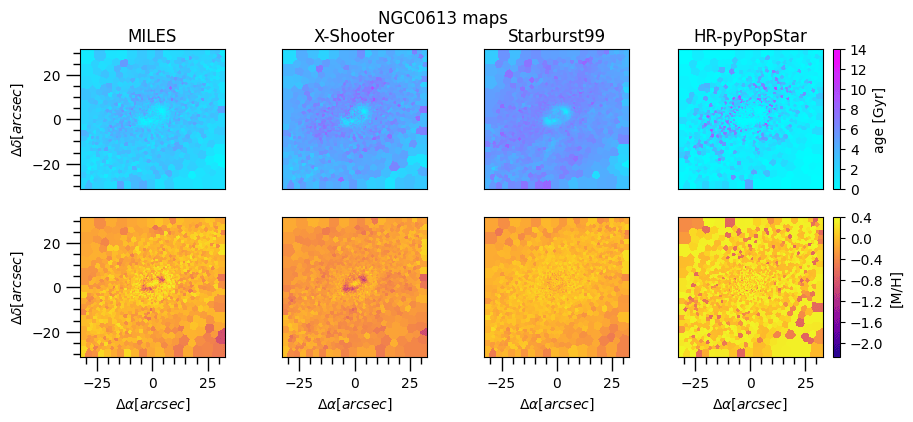}
	\caption{Same as Figure \ref{NGC3351_maps}, but for NGC\,0613.}
	\label{NGC0613_maps}
\end{figure*}

\begin{figure*}
	\centering
	\includegraphics[width=\columnwidth]{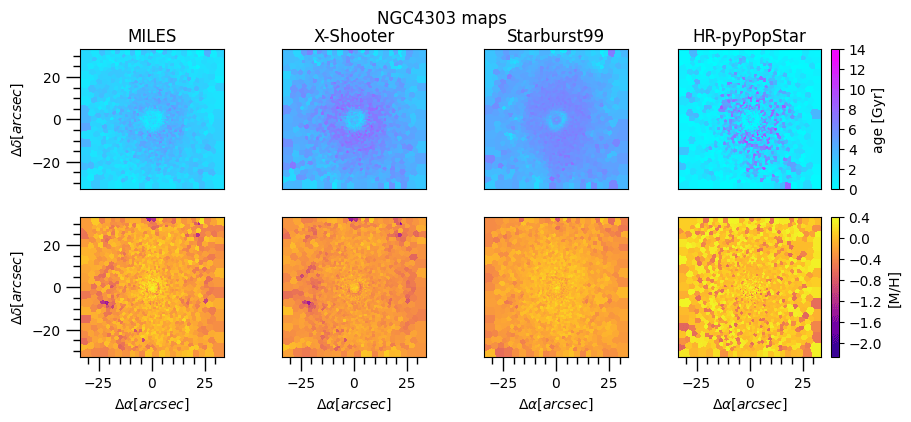}
	\caption{Same as Figure \ref{NGC3351_maps}, but for NGC\,4303.}
	\label{NGC4303_maps}
\end{figure*}

\begin{figure*}
	\centering
	\includegraphics[width=\columnwidth]{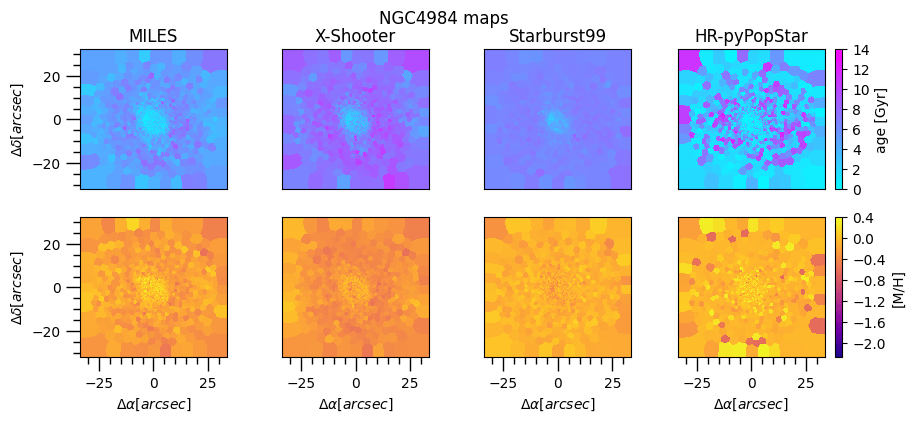}
	\caption{Same as Figure \ref{NGC3351_maps}, but for NGC\,4984.}
	\label{NGC4984_maps}
\end{figure*}

\begin{figure*}
	\centering
	\includegraphics[width=\columnwidth]{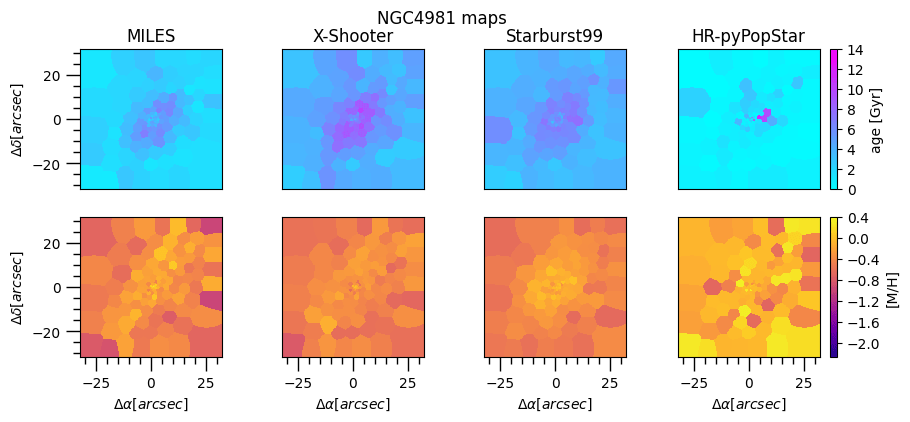}
	\caption{Same as Figure \ref{NGC3351_maps}, but for NGC\,4981.}
	\label{NGC4981_maps}
\end{figure*}

\begin{figure*}
	\centering
	\includegraphics[width=\columnwidth]{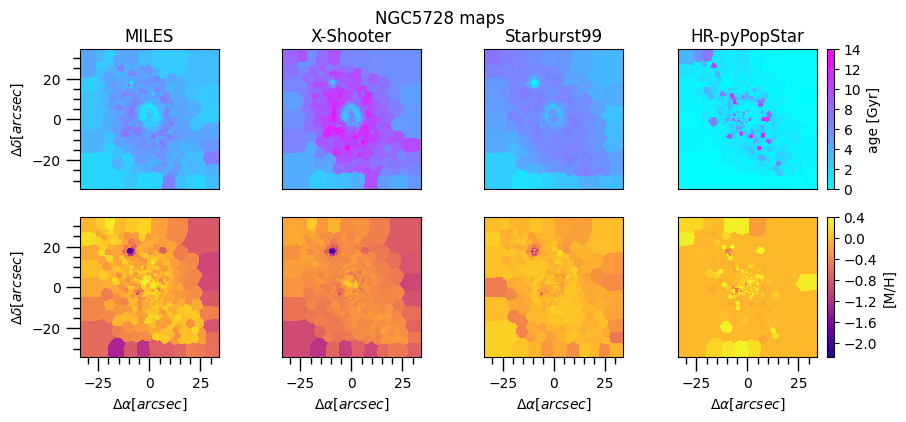}
	\caption{Same as Figure \ref{NGC3351_maps}, but for NGC\,5728.}
	\label{NGC5728_maps}
\end{figure*}

\begin{figure*}
	\centering
	\includegraphics[width=\columnwidth]{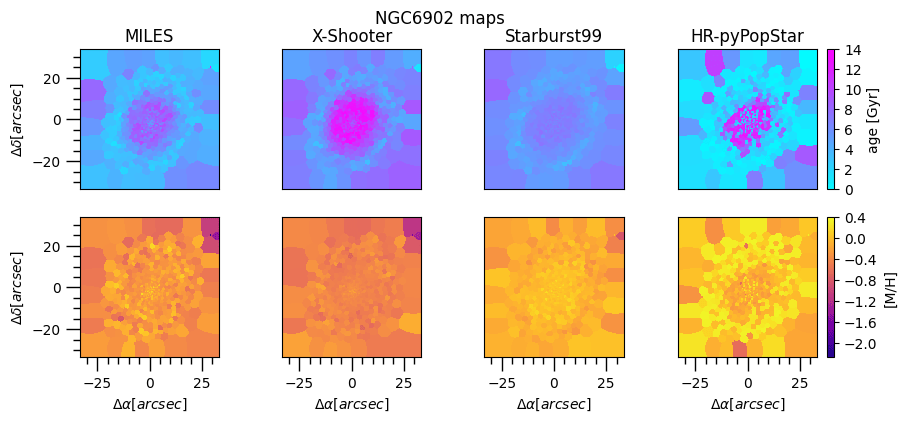}
	\caption{Same as Figure \ref{NGC3351_maps}, but for NGC\,6902.}
	\label{NGC6902_maps}
\end{figure*}

\subsection{Emission lines}

Figure \ref{gas_metal_maps} shows the gas-phase metallicity maps for the whole sample of galaxies in this work.

\begin{figure*}
	\centering
	\includegraphics[width=0.85\columnwidth]{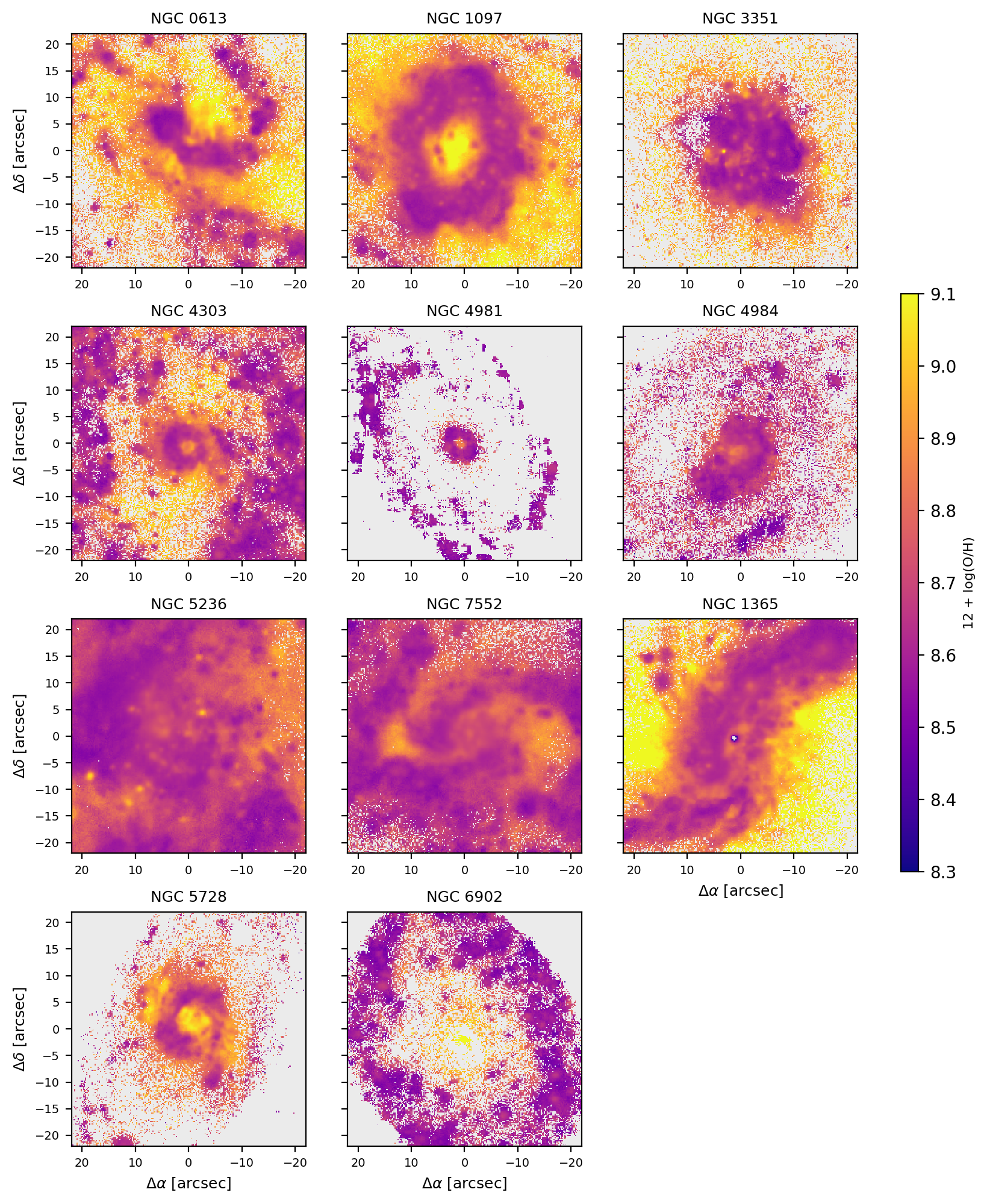}
	\caption{\textbf{Gas-phase metallicity maps derived from the S-calibration for the full sample of 11 TIMER galaxies, on the same colour scale and angular scale as Fig.~\ref{gas_metal_maps_1}, with the $1\sigma$ selection applied. Most galaxies show lower gas metallicities in their central star-forming regions than in their surroundings, qualitatively consistent with bar-driven inflows of chemically diluted gas. The abundance is close to solar in all of them, and the variation across each field is a few tenths of a dex. We caution that the amplitude of any such contrast depends on the adopted signal-to-noise threshold; all maps here use the $1\sigma$ selection.}}
	\label{gas_metal_maps}
\end{figure*}

\bsp	%
\label{lastpage}
\end{document}